\documentclass[pre,superscriptaddress,twocolumn]{revtex4}
\usepackage{graphicx}
\usepackage{bm}
\usepackage{amsmath}
\usepackage{amssymb}
\usepackage{balance}
\usepackage{exscale}
\usepackage{wasysym}
\usepackage{ulem}
\usepackage{cancel}
\bibstyle{apsrev.bib}

\usepackage{xcolor}

\newcommand{\vect}[1]{\mathbf{{#1}}}
\newcommand{\Ham}{\mathcal{H}}
\begin{document}

\title{Generalized Langevin equation with shear flow and its fluctuation-dissipation theorems derived from  a Caldeira-Leggett Hamiltonian}

\author{Sara Pelargonio}

\email[Electronic mail: \ ]{sara.pelargonio@studenti.unimi.it}

\affiliation{Department of Physics ``A. Pontremoli", University of Milan, via Celoria 16, 20133 Milan, Italy}

\author{Alessio Zaccone}

\email[Electronic mail: \ ]{alessio.zaccone@unimi.it}

\affiliation{Department of Physics ``A. Pontremoli", University of Milan, via Celoria 16, 20133 Milan, Italy}


\date{\today}

\begin{abstract}
We provide a first-principles derivation of the Langevin equation with shear flow and its corresponding fluctuation-dissipation theorems. Shear flow of simple fluids has been widely investigated by numerical simulations. Most studies postulate a Markovian Langevin equation with a simple shear drag term à la Stokes. However, this choice has never been justified from first principles. We start from a particle-bath system described by a classical Caldeira-Leggett Hamiltonian modified by adding a term proportional to the strain-rate tensor according to Hoover's DOLLS method, and we derive a generalized Langevin equation for the sheared system. We then compute, analytically, the noise time-correlation functions in different regimes. Based on the intensity of the shear-rate, we can distinguish between close-to-equilibrium and far-from-equilibrium states. According to the results presented here, the standard, simple and Markovian form of the Langevin equation with shear flow postulated in the literature is valid only in the limit of extremely weak shear rates compared to the effective vibrational temperature of the bath. For even marginally higher shear rates, the (generalized) Langevin equation is strongly non-Markovian and non-trivial fluctuation-dissipation theorems are derived.
\end{abstract}

\maketitle

\section{Introduction}
 The theory of Brownian motion is a milestone of nonequilibrium statistical mechanics as it provides a simple and valid approximation 
 to the dynamics of nonequilibrium systems. The motion of a Brownian particle interacting with a heat-bath of solvent molecules and under the effect of external force fields can be described either by a Langevin \cite{Graham,cui2018generalized} or by a Smoluchowski (Fokker-Planck) equation \cite{dhont1996introduction,Banetta}. In the Langevin's approach, the interaction of a particle with  the solvent (hidden) molecules produces both friction and randomness: the rapid changes in the particle's velocity are dissipated by viscosity. The Langevin equation is at the basis of diffusion equations and fluctuation-dissipation theorems. In fact, as noise and friction have the same physical origin in the interaction with heat-bath's molecules, there is a mathematical relation that links the noise fluctuations to the friction. Traditionally, this relation follows the second fluctuation-dissipation theorem (FDT) and gives a proportionality between the noise's fluctuations amplitude and the friction kernel \citep{van1992stochastic, zwanzig2001nonequilibrium}
 
 \begin{equation}
     \langle F(t) F(s) \rangle \propto k_B T K(t-s).
 \end{equation}
 
Usually, when dealing with Langevin equations, some hypothesis on the average of the noise and on the expression of the friction kernel are assumed. In most of the applications the noise function is assumed to be "white", that is, it has zero mean and a time-correlation function  proportional to a Dirac delta. Langevin equations with white noise and constant friction coefficient are used to describe Markov processes.
 
The Brownian motion has been widely studied in the case of particles in solvents which are not perturbed by any external driving force \citep{kubo1986brownian,zwanzig2001nonequilibrium}. 
 However, many studies extended its applications to nonequilibrium and driven solvents \citep{PhysRevE.94.062139, imparato2007work} such as fluid flows. 
 Among fluid flows, shear flow is one of the most common examples for its theoretical and practical applications such as in aerospace engineering \cite{fiedler1990management, smith1997vectoring}, atmospheric physics   \cite{falkovich2002acceleration, huang1979theory}, colloidal systems \cite{dhont1996introduction,Zaccone_2009,Tsoutsoura} or plasma physics \cite{doi:10.1081/SMTS-200056102, tynan2006observation}. It is important also in the study of soft matter systems to determine the microstructure of colloidal dispersions under flow conditions \citep{wagner2009shear, gurnon2015microstructure,Banetta,Riva,Anzivino} and to understand and describe diffusion of solutes in channel flows \cite{Taylor}.

The Langevin formalism is widely used in  numerical simulations to study the physical properties of nonequilibrium glassy materials \citep{berthier2002nonequilibrium,berthier2002shearing, PhysRevLett.124.225502}. Usually, when performing simulations, a Langevin equation with shear is postulated \citep{muthu,ikeda2012unified, derksen1990light}, both in its original underdamped form \cite{McPhie}

\begin{equation} \label{lang_simultions_1}
    m \dot{\vect{v}} = -\zeta\vect{p} + \vect{F}(t) - V'[\vect{r}] + \dot{\gamma} \zeta y\hat{\vect{x}}
\end{equation}

\noindent  and in the overdamped limit \citep{berthier2002shearing, fuchs2009nonlinear} 

\begin{equation}\label{lang_simultions_2}
  \zeta \bigl( \vect{v}(t) - \dot{\gamma}y \hat{\vect{x}}\bigr) = -V'[\vect{r}] + \vect{F}(t)  
\end{equation}

\noindent where $\vect{r}$ is the particle's position, $\zeta$ is the friction coefficient, $y$ is the particle's position component along the $y$ axis, and $\dot{\gamma}$ is the strain-rate or shear-rate.

However, in spite of its wide use, a first-principles mathematical derivation of the Langevin equation with shear flow, including the form of the shear-rate dependent friction kernel,  has apparently not been provided yet.
Among previous attempts, we shall mention the important work of Ref. \cite{McPhie}, where the Langevin equation with shear flow was derived from first principles using the Mori-Zwanzig projection operator method. However, no explicit form of the friction memory kernel was obtained. Without the knowledge of the memory kernel, it is impossible to study the stochastic properties of the noise and to derive meaningful fluctuation-dissipation relations.

Shear-flow systems have been studied as examples of driven systems to investigate the validity of fluctuation-dissipation relations out of equilibrium \citep{speck2009extended, seifert2010fluctuation, berthier2002nonequilibrium}. It was found that the first fluctuation-dissipation theorem (FDT) is different from the equilibrium case \cite{sarracino2019fluctuation, baldovin2022many,Rubi, seifert2010fluctuation} but, on the other hand, according to some recent studies \cite{D1SM00521A,Jung_2022}, the second FDT should be valid even far from equilibrium. For this reason, the second FDT allows one to study some properties of a system even in more general cases such as far from equilibrium. 

Some studies \citep{PhysRevE.91.022128, maes2014second} proved that, in the case of driven environments, an extension of the second FDT is needed. For this reason, it would be relevant to derive a second FDT for systems under shear flow from first principles. It is known, in fact, that most of the times the Markovianity is just an approximation of a real non-Markovian behaviour, and it is still not clear whether, and, eventually, how, external force fields may affect the stochastic nature of a sheared system. Therefore, it would be relevant to provide a derivation of an extended FDT for shear-flow systems, without imposing any a priori assumption on the nature of the stochastic noise. 

We therefore provide a rigorous derivation of a Langevin equation and its related second FDT with shear flow from first principles. We proceed by following the particle-bath Hamiltonian approach as it was at first proposed by Zwanzig \citep{zwanzig2001nonequilibrium} to derive a Langevin equation for simple Brownian motion. This method has already been used in other cases of externally time-dependent driven systems \citep{cui2018generalized} but it has never been applied to the case of fluids under external shear forces, which is paradigmatic for a whole class of nonequilibrium dissipative systems.

\section{Theory}

\subsection{DOLLS tensor method}
Historically, the first method developed for performing calculations on viscous flows was Hoover's DOLLS tensor method \citep{PhysRevA.22.1690}. It is based on the idea that any mechanical flow can be described by specifying the space and time dependence of the strain rate tensor $\nabla\vect{u}$, which describes the rate at which any internal coordinate $\vect{q}$ changes with time according to

\begin{equation}
\label{eqn:DOLLS}
\dot{\vect{q}} = \vect{q}\cdot{\nabla\vect{u}}. 
\end{equation}

\noindent   If standard conservation principles are valid, then the corresponding change in momentum is: 

\begin{equation}
     \dot{\vect{p}} = -\nabla\vect{u}\cdot{\vect{p}}
\end{equation}

A microscopic Hamiltonian which contains the usual changes in coordinates and momenta from potential and kinetic energy terms as well as changes due to an applied macroscopic mechanical deformation described by the DOLLS tensor $\vect{q}\vect{p} : \nabla\vect{u}$ reads as:
 
 \begin{equation}
     \Ham = \Phi(\vect{q}) + K(\vect{p}) + \vect{q}\vect{p} : \nabla\vect{u}.
 \end{equation}

The corresponding equations of motion can be integrated numerically and used to perform numerical simulations on fluid flow systems  \cite{hoover1983atomistic}.

Later, an alternative computational approach, called SLLOD, was developed in \citep{book,EVANS1984297,PhysRevA.30.1528}. It is based on different microscopic equations of motion that still describe the same macroscopic flow and are consistent with thermodynamics principles. However, unlike the DOLLS tensor equations, the SLLOD equations of motion cannot be derived from any Hamiltonian. The SLLOD equations of motion read as follows:
\begin{equation}
\begin{split}
 &\dot{\vect{q}}= \frac{\vect{p}}{m} + \vect{q}\cdot \nabla \vect{u} \\
&\dot{\vect{p}} = \vect{F}-\vect{p}\cdot \nabla\vect{u} .  
\end{split}
\end{equation}

It has been demonstrated  \citep{book}  that these equations give the same dissipation as the DOLLS equations. Many studies \citep{PhysRevE.78.046701, baig2005proper, evans1984nonlinear,hoover2009nonlinear, EVANS1984297} compared the results obtained by DOLLS and SLLOD algorithms pointing out that, in some instances, they may lead to different results, such as the direction of the rotation of particles under shear, or slightly different predicted values for the normal stress. Even though the SLLOD method is usually preferred, the DOLLS tensor Hamiltonian is still widely adopted as the only Hamiltonian describing a microscopic flow, and for this reason it will taken as the starting point of the model adopted here.

\subsection{The model}
\begin{figure}
    \includegraphics[scale = 0.35]{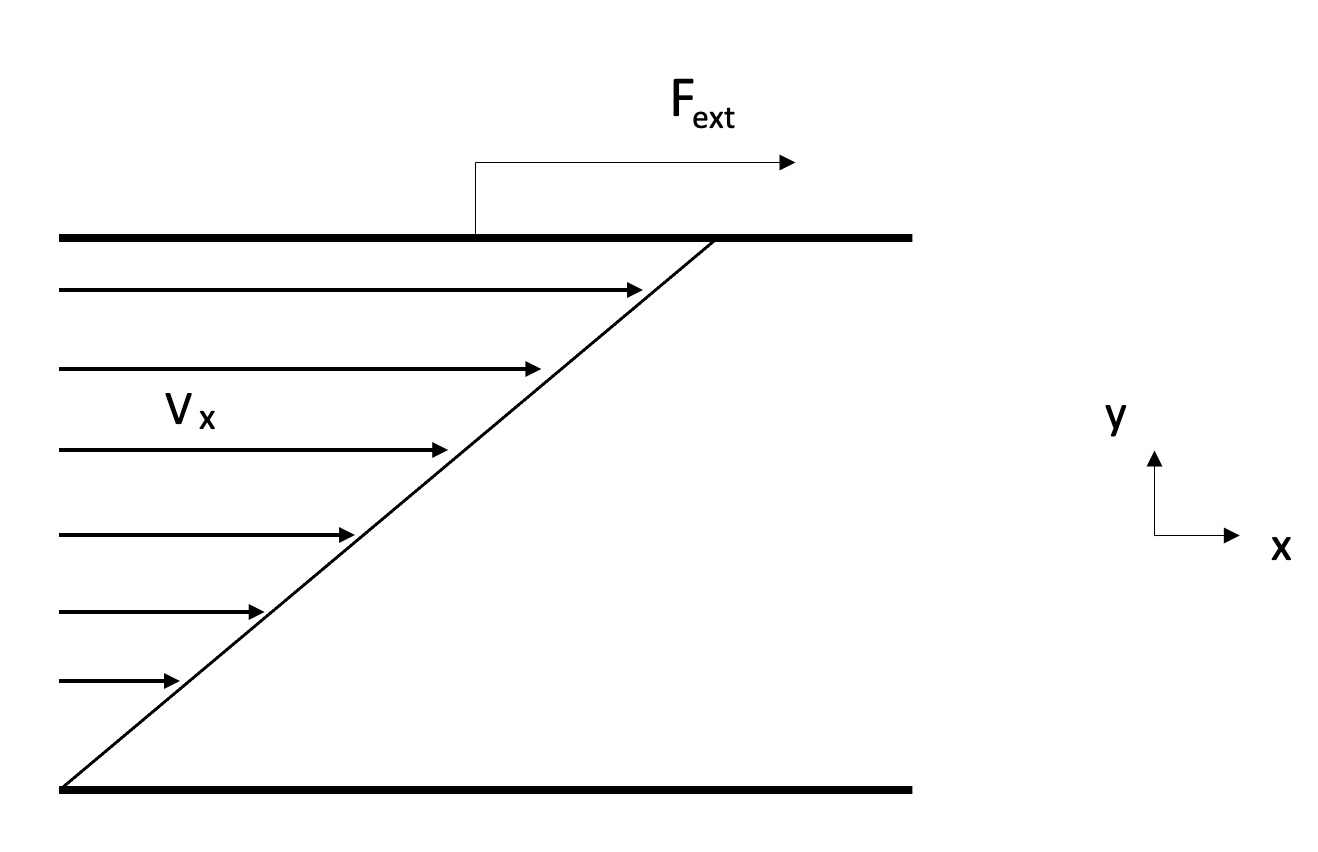}
    \caption{Schematic of a planar Couette simple shear flow used in the derivations.}
    \label{fig:Couette}
\end{figure}

We consider a homogeneous planar Couette flow, cfr. Fig. \ref{fig:Couette}, described by the strain-rate tensor 

\begin{equation}\label{strain_tensor}
\nabla \vect{u} = \begin{pmatrix} \frac{\partial u_x}{\partial x} &  \frac{\partial u_y}{\partial x} \\  \frac{\partial u_x}{\partial y} &  \frac{\partial u_y}{\partial y} \end{pmatrix} = \begin{pmatrix} 0 & 0 \\ \dot{\gamma} & 0 \end{pmatrix}
\end{equation}

\noindent
where $\dot{\gamma}$ is the shear rate that, for simplicity, is assumed to be constant.  The strain-rate tensor determines a change in velocity equal to 

\begin{equation}
 \vect{v} =  \nabla\vect{u} \cdot \vect{x} = \dot{\gamma} y \hat{\vect{x}} .
\end{equation}

 We  study the classical version of the Caldeira-Leggett coupling \citep{CALDEIRA1983374} between a tagged particle and a bath of harmonic oscillators, as already proposed by Zwanzig, and add a new term that describes the microscopic deformation that determines the fluid flow. As discussed above, this term can be taken to be proportional to the strain-rate tensor and to the DOLLS tensor. We therefore take the Hamiltonian of the system of interest, namely the colloidal particle, and of its environment, that are the bath's degrees of freedom, to be 

\begin{equation}
\begin{split}
 &\Ham_{P} = \frac{P^{2}}{2M} + V(\vect{Q}) + \vect{Q}\vect{P}:\nabla\vect{u} \\
 &\Ham_{B} = \sum_{i} \Bigl[ \frac{p_{i}^{2}}{2 } + \frac{1}{2}  \omega_{i}^{2} \Bigl(\vect{q}_{i} - \frac{c_{i}}{ \omega_{i}^{2}} \vect{Q} \Bigr )^{2} + \vect{q}_i\vect{p}_i:\nabla\vect{u} \Bigr]
\end{split}
\end{equation}

\noindent where $(\vect{P},\vect{Q})$ are the tagged particle's coordinates, while $(\vect{p}_i,\vect{q}_i)$ are the heat bath's degrees of freedom. They are modelled as harmonic oscillators with oscillation frequency $    \omega_i$ and interact bilinearly with the tagged particle. The tagged particle is assumed to perturb its environment weakly, so the coupling is assumed to be small. The coupling's strength is determined by the constant $c_i$, known as the strength of coupling between the tagged particle and the $i$-th bath oscillator. By using the DOLLS strain-rate tensor Eq. \eqref{strain_tensor}, the Hamiltonians become
\begin{equation} \label{hamiltonian}
\begin{split}
 &\Ham_{P} = \frac{P^{2}}{2M} + V(\vect{Q}) +\dot{\gamma} Q_y P_x \hat{\vect{x}}\hat{\vect{y}}\\
& \Ham_{B} = \sum_{i}\Bigl[ \frac{p_{i}^{2}}{2 } + \frac{1}{2}  \omega_{i}^{2} \Bigl(\vect{q}_{i} - \frac{c_{i}}{ \omega_{i}^{2}} \vect{Q} \Bigr )^{2} + \dot{\gamma} q_{iy} p_{ix}\hat{\vect{x}}\hat{\vect{y}}\Bigr].
\end{split}
\end{equation}

Apart from the terms in $\dot{\gamma}$, which are new and stem from the DOLLS model for the strain rate applied to the system, the above Hamiltonian coincides with the standard (classical) Caldeira-Leggett Hamiltonian used throughout the literature \cite{zwanzig2001nonequilibrium,CALDEIRA1983374,Vacchini,cui2018generalized,Gottwald,Ceotto}.

The second term in the bracket in the definition of $\Ham_{B}$ is manifestly breaking translation invariance. It should be noted that a suitably renormalized potential $\tilde{V}(\vect{Q})\equiv V(\vect{Q})-\frac{1}{2}\sum_{i}\frac{c_{i}^{2}}{\omega_{i}^{2}}Q^{2}$, provides a ``counter-term'' that makes the Hamiltonian translation-invariant, as emphasized by various authors \cite{CALDEIRA1983374,Vacchini,Weiss,Gottwald}.
Since in our study the potential $V(\vect{Q})$ is left unspecified, we can implicitly assume that our $V(\vect{Q})$ in Eq. \eqref{hamiltonian} contains such counter-term so that the Hamiltonian is translation-invariant.

Furthermore, it is assumed that the external deformation field acts both on the tagged particle and on the heat bath's particles and all the results presented in the next sections are derived from this hypothesis. It is not the only possible case, as one could expect the tagged particle to flow also because of the interactions with the heat-bath's degrees of freedom, with only the latter being subjected to the deformation. We shall demonstrate later on that this hypothesis leads to a slightly different form of the Langevin equation with shear.
 \\

\subsection{Derivation}
The Hamiltonian Eq. \eqref{hamiltonian} leads to the following system of equations of motion

\begin{equation} \label{ODEs}
\begin{split}
&\dot{Q}_x = \frac{P_x}{M}+\dot{\gamma}Q_y(t) \\
&\dot{Q}_y =\frac{P_y}{M} \\
&\dot{P}_x =-V'[\vect{Q}]_x  +\sum_i c_i\Bigl(q_{ix}-\frac{c_i Q_x}{\omega_i^2}\Bigr) \\
&\dot{P}_y = -V'[\vect{Q}]_y - \dot{\gamma}P_y(t) +\sum_i c_i\Bigl(q_{iy}-\frac{c_i Q_y}{\omega_i^2}\Bigr)\\
&\dot{q}_{ix} = p_{ix} + \dot{\gamma} q_{iy}   \\
&\dot{q}_{iy} = p_{iy} \\
&\dot{p}_{ix} = -\omega^2 q_{ix} + c_i Q_x \\
&\dot{p}_{iy} = -\omega^2 q_{iy} - \dot{\gamma} p_{ix} + c_i Q_y.
\end{split}
\end{equation}

We first consider the ODE system for the bath's coordinates. It can be solved by considering the vectors' components separately. In matrix form, it reads 
\begin{equation}
\dot{\vect{y}} = \vect{A}\vect{y} + \vect{C}
\end{equation}

\noindent 
where

\begin{equation}
\vect{y} = \begin{pmatrix} q_x \\ q_y \\ p_x \\ p_y
\end{pmatrix}
\quad 
\vect{ A} =  \begin{pmatrix} 0 & \dot{\gamma} & 1 & 0 \\
 0 & 0 & 0 & 1 \\
 -\omega^2 & 0 & 0 & 0 \\
 0 & -\omega^2 & -\dot{\gamma} & 0 
 \end{pmatrix}
 \quad
 \vect{C} = \begin{pmatrix} 0 \\ 0 \\ Q_x \\ Q_y 
 \end{pmatrix}
 \end{equation}

\noindent and is solved by diagonalizing $\vect{A}$, changing variables to $\vect{A}$'s eigenvectors basis, solving the decoupled system using Duhamel's formula, and then performing the inverse change of variables. It leads to

\begin{widetext} 
\begin{equation} \label{sol_bath_x}
\begin{split} 
q_{ix}(t) &= \frac{c_i}{4\omega_i} \int_0^t dt' Q_y(t') \Bigl(e^{-\lambda_2(t-t')} + e^{\lambda_2(t-t')}\Bigr) 
+ \frac{c_{i}}{4\omega_i} \frac{\sqrt{\dot{\gamma}-\omega_i}}{\sqrt{\omega_i}} \int_0^t dt' Q_x(t') \Bigl(e^{-\lambda_2(t-t')}- e^{\lambda_2(t-t')}\Bigr) \\
&-\frac{c_i}{4\omega_i} \int_0^t dt' Q_y(t') \Bigl(e^{\lambda_4(t-t')}+ e^{-\lambda_4(t-t')}\Bigr)-\frac{c_i}{4\omega_i} \frac{\sqrt{\dot{\gamma}+\omega_i}}{\sqrt{-\omega_i}}  \int_0^t dt' Q_x(t') \Bigl(e^{-\lambda_4(t-t')}- e^{\lambda_4(t-t')}\Bigr)\\
& 
+ \frac{\sqrt{\dot{\gamma}-\omega_i}}{4\omega_i^2 \sqrt{\omega_i}}p_{ix}(0)\Bigl(e^{-\lambda_2 t}-e^{\lambda_2 t}\Bigr)+\frac{p_{iy}(0)}{4\omega_i}\Bigl(e^{-\lambda_2 t}+e^{\lambda_2t}\Bigr)
-\frac{\omega_i\sqrt{-\omega_i(\dot{\gamma}+\omega_i)}}{4\omega_i^2}p_{ix}(0) e^{\lambda_4 t} \\ &-\frac{p_{iy}(0)}{4\omega_i}\Bigl(e^{-\lambda_4 t}+e^{\lambda_4t}\Bigr)-\frac{\dot{\gamma}+\omega_i}{4\omega_i\sqrt{-\omega_i(\dot{\gamma}+\omega_i)}} p_{ix}(0)e^{-\lambda_4t} +\frac{\sqrt{-\omega_i(\dot{\gamma}+\omega_i)}}{4\omega_i}q_{iy}(0)\Bigl(e^{\lambda_2 t}-e^{-\lambda_2t}\Bigr)\\& +\frac{\sqrt{\omega_i(\dot{\gamma}-\omega_i)}}{4\omega_i}q_{iy}(0)\Bigl(e^{-\lambda_4 t}-e^{\lambda_4t}\Bigr)+\frac{q_{ix}(0)}{4}\Bigl(e^{-\lambda_4 t}+e^{\lambda_4t}\Bigr) +\frac{q_{ix}(0)}{4}\Bigl(e^{-\lambda_2 t}+e^{\lambda_2t}\Bigr)  \end{split}
\end{equation}

\noindent and

\begin{equation} \label{sol_bath_y}
\begin{split}
q_{iy}(t) =& \frac{c_i}{4\sqrt{\omega_i(\dot{\gamma}-\omega_i)}} \int_0^t dt' Q_y(t') \Bigl( e^{\lambda_2(t-t')}-e^{-\lambda_2(t-t')}\Bigr) - \frac{c_i}{4\omega_i} \int_0^t dt' Q_x(t') \Bigl( e^{\lambda_2(t-t')}+e^{-\lambda_2(t-t')}\Bigr) \\
&+ \frac{c_i}{4\sqrt{-\omega(\dot{\gamma}+\omega_i)}} \int_0^t dt' Q_y(t') \Bigl( e^{\lambda_4(t-t')}-e^{-\lambda_4(t-t')}\Bigr) +\frac{c_i}{4\omega_i} \int_0^t dt' Q_x(t') \Bigl( e^{\lambda_4(t-t')}+e^{-\lambda_4(t-t')}\Bigr) \\ 
&+\frac{p_{ix}(0)}{4\omega_i}\Bigl(e^{\lambda_2 t}+e^{-\lambda_2 t}\Bigr)-\frac{p_{ix}(0)}{4\omega_i}\Bigl(e^{\lambda_4 t}+e^{-\lambda_4 t}\Bigr)  +\frac{p_{iy}(0)}{4\sqrt{\omega_i(\dot{\gamma}-\omega_i)}}\Bigl(e^{\lambda_2 t}-e^{-\lambda_2 t}\Bigr)\\
&+\frac{p_{iy}(0)}{4\sqrt{-\omega_i(\dot{\gamma}+\omega_i)}}\Bigl(e^{\lambda_4 t}-e^{-\lambda_4 t}\Bigr)+\frac{\omega_i}{4\sqrt{\omega_i(\dot{\gamma}-\omega_i)}}q_{ix}(0)\Bigl(e^{\lambda_2 t}-e^{-\lambda_2 t}\Bigr)\\
&-\frac{\omega_i}{4\sqrt{-\omega_i(\dot{\gamma}+\omega_i)}}q_{ix}(0)\Bigl(e^{\lambda_4 t}-e^{-\lambda_4 t}\Bigr)
+\frac{q_{iy}(0)}{4}\Bigl(e^{\lambda_2 t}+e^{-\lambda_2 t}\Bigr) +\frac{q_{iy}(0)}{4}\Bigl(e^{\lambda_4 t}+e^{-\lambda_4t}\Bigr)\\
\end{split}
\end{equation}
\end{widetext}

\noindent where

\begin{equation}
    \begin{split}
\label{eigenvalues}
    \lambda_2 = \sqrt{\omega_i(\dot{\gamma}-\omega_i)} \\ \lambda_4 = \sqrt{-\omega_i(\dot{\gamma}+\omega_i)}
\end{split}
\end{equation}

 \noindent are \textbf{A}'s eigenvalues along with $ \lambda_1 = -\lambda_2$ and $\lambda_3 = -\lambda_4$.
 Upon evaluating the integrals by parts and plugging these solutions into the equations for $P_x(t)$ and $P_y(t)$ in Eq. \eqref{ODEs}, we find the following generalized Langevin equation (GLE) with shear flow: 

\begin{widetext}
\begin{equation}  \label{lang_gen}
\begin{split}
    \dot{\vect{P}} &= -V'[\vect{Q}] -\dot{\gamma}P_x(t)  + \vect{F}(t)  - \int_0^t dt' \vect{K}_{tot}(t-t') \dot{\vect{Q}}(t') 
\end{split}
\end{equation}
Terms explicitly depending on $\vect{Q}$ or one of its Cartesian components, are not allowed into the equation of motion because
they depend on the position of the particle, and therefore have to vanish for a system with translational invariance, as noted already by Andersen \cite{Andersen} and by Ray and Rahman \cite{Rahman}.

\end{widetext}

Furthermore, the memory kernel is the sum of four terms: $\vect{K}_{tot}=\vect{K}_{1}+\vect{K}_{2}+\vect{K}_{3}+\vect{K}_{4}$ which are given by the following expressions:

\begin{widetext}

\begin{equation}
\begin{split}
   & \vect{K}_1 = \begin{pmatrix}
   0 & -\sum_i \frac{c_i^2}{4\omega_i \sqrt{\omega_i(\dot{\gamma}-\omega_i)}}  \Bigl(e^{\lambda_2(t-t')} - e^{-\lambda_2(t-t')} \Bigr)\\  \sum_i \frac{c_i^2}{4\omega_i \sqrt{\omega_i(\dot{\gamma}-\omega_i)}}\Bigl(e^{\lambda_2(t-t')} - e^{-\lambda_2(t-t')} \Bigr) & 0 
   \end{pmatrix}  \\
   &    \vect{K}_2 = \begin{pmatrix}  \sum_i \frac{c_i^2}{4\omega_i^2} \Bigl(e^{\lambda_2(t-t')} + e^{-\lambda_2(t-t')} \Bigr)  & 0 \\ 0 & -\sum_i \frac{c_i^2}{4\omega_i (\dot{\gamma}-\omega_i)} \Bigl(e^{\lambda_2(t-t')} + e^{-\lambda_2(t-t')} \Bigr)
    \end{pmatrix} \\
    & \vect{K}_3 = \begin{pmatrix}
    0 &  \sum_i\frac{c_i^2}{4\omega_i \sqrt{-\omega_i(\dot{\gamma}+\omega_i)}} \Bigl(e^{\lambda_4(t-t')} + e^{-\lambda_4(t-t')} \Bigr) \\ -\sum_i \frac{c_i^2}{4\omega_i \sqrt{-\omega_i(\dot{\gamma}+\omega_i)}} \Bigl(e^{\lambda_4(t-t')} - e^{-\lambda_4(t-t')} \Bigr) & 0 
   \end{pmatrix}  \\
    &    \vect{K}_4 = \begin{pmatrix} \sum_i \frac{c_i^2}{4\omega_i^2} \Bigl(e^{\lambda_4(t-t')} + e^{-\lambda_4(t-t')} \Bigr)& 0 \\ 0 & \sum_i \frac{c_i^2}{4\omega_i (\dot{\gamma}+\omega_i)} \Bigl(e^{\lambda_4(t-t')} + e^{-\lambda_4(t-t')} \Bigr) 
    \end{pmatrix} \end{split}
\end{equation}

\end{widetext}

\noindent and the noise function is 

\begin{equation}
 \vect{F} (t) = \begin{pmatrix} F_x (t) \\ F_y (t) \end{pmatrix}
\end{equation}

\noindent with

\begin{widetext}
\begin{equation} \label{noise_x}
\begin{split}
F_{x}(t) &=  \sum_i\Biggl\{
\frac{c_i\sqrt{\dot{\gamma}-\omega_i}}{4\omega_i\sqrt{\omega_i}}p_{ix}(0) \Bigl(e^{-\lambda_2 t}-e^{\lambda_2 t}\Bigr)
+\Bigl[-\frac{c_i^2}{\omega_i^2}Q_x(0)+\frac{c_i}{4}q_{ix}(0)+\frac{c_i}{4\omega_i} p_y(0)\Bigr]\Bigl(e^{\lambda_2 t}+e^{-\lambda_2 t}\Bigr)\\&
+ \Bigl[-\frac{c_i^2}{4\omega_i^2}Q_x(0) + \frac{c_i}{4} q_{ix}(0) - \frac{c_i}{4\omega_i} p_{iy}(0)\Bigr]\Bigl(e^{\lambda_4t} + e^{-\lambda_4 t}\Bigr) -\frac{c_i}{4\omega_i^2}\sqrt{-\omega_i(\dot{\gamma}+\omega_i)} p_{ix}(0)e^{\lambda_4 t}\\&+\Biggl[ \frac{c_i^2}{4\omega_i\sqrt{\omega_i(\dot{\gamma}-\omega_i)}}Q_y(0) - \frac{c_i\sqrt{\omega_i(\dot{\gamma}-\omega_i})}{4\omega_i}q_{iy}(0)\Biggr]\Bigr(e^{\lambda_2 t}-e^{-\lambda_2 t}\Bigr) 
\\&+ \Biggl[ -\frac{c_i^2}{4\omega_i\sqrt{-\omega_i(\dot{\gamma}+\omega_i)}}Q_y(0)+  \frac{c_i\sqrt{-\omega_i(\dot{\gamma}+\omega_i})}{\omega_i}q_{iy}(0)\Biggr] \Bigr(e^{\lambda_4 t}-e^{-\lambda_4 t}\Bigr) \\
&- c_i\frac{\dot{\gamma}+\omega_i}{4\omega_i\sqrt{-\omega_i(\dot{\gamma}+\omega_i)}} p_{ix}(0) e^{-\lambda_4 t}
  \Biggr\}
\end{split}
\end{equation}

\begin{equation} \label{noise_y}
\begin{split}
F_y (t) &= \sum_i \Biggl\{
\Biggl(-\frac{c_i^2 Q_y(0)}{4\omega_i(\dot{\gamma}-\omega_i)} + \frac{c_i}{4} q_{iy}(0) -\frac{c_i}{4\omega_i} p_{ix}\Biggr) \Bigl(e^{\lambda_2t'
}+e^{-\lambda_2 t'} \Bigr)  \\
&+ \Biggl( \frac{c_i^2 Q_x(0)}{4\omega_i\sqrt{-\omega_i(\dot{\gamma}+\omega_i)}}-\frac{c_i\omega_i}{\sqrt{-\omega_i(\dot{\gamma}+\omega_i)}}q_{ix}(0)+\frac{c_i p_{iy}(0)}{4\sqrt{-\omega_i(\dot{\gamma}+\omega_i)}}
\Biggr) \Bigl(e^{\lambda_4 t'}-e^{-\lambda_4 t'}\Bigr)\\& 
+ \Biggl(-\frac{c_i^2 Q_x(0)}{4\omega_i\sqrt{\omega_i(\dot{\gamma}-\omega_i)}}+\frac{c_i\omega_i}{\sqrt{\omega_i(\dot{\gamma}-\omega_i)}}q_{ix}(0) + \frac{c_i p_{iy}(0)}{4\sqrt{\omega_i(\dot{\gamma}-\omega_i)}} \Biggr)
\Bigl(e^{\lambda_2 t'}-e^{-\lambda_2 t'}\Bigr) \\ &+ \Biggl(-\frac{c_i^2 Q_y(0)}{4\omega_i(\dot{\gamma}+\omega_i)} + \frac{c_i}{4} q_{iy}(0)
+\frac{c_i}{4\omega_i} p_{ix}\Biggr)\Bigl(e^{\lambda_4t'}+e^{-\lambda_4 t'} \Bigr)\Biggr\}
    \end{split}
\end{equation}

\end{widetext}

We then compute the noise time-correlation function assuming the heat bath's initial conditions to be drawn from a Boltzmann distribution \citep{zwanzig2001nonequilibrium}

\begin{equation}
    \langle \vect{F}(t) \vect{F}(t') \rangle = \int_{-\infty}^{+\infty} d\vect{q}(0) d\vect{p}(0) \vect{F}(t) \vect{F}(t') e^{-\Ham_B/k_B T}. 
\end{equation}

Depending on when the external shear deformation begins to act on the system, there are two different physical situations. We can consider a system that is in mechanical equilibrium at $t=0$ with the external shear perturbation acting from $t>0$ or a system that is already sheared, and not in equilibrium, at $t=0$. That is, we can compute two different time-correlation functions, leading to two different fluctuation-dissipation relations, by taking the Boltzmann's weight $\Ham_B$ to be 

\begin{equation} \label{ham_noshear}
    \Ham_{B} = \sum_{i}\Bigl[ \frac{p_{i}^{2}}{2 } + \frac{1}{2}  \omega_{i}^{2} \Bigl(\vect{q}_{i} - \frac{c_{i}}{ \omega_{i}^{2}} \vect{Q} \Bigr )^{2} \Bigr]
\end{equation}
or
\begin{equation}
    \Ham_{B} = \sum_{i}\Bigl[ \frac{p_{i}^{2}}{2 } + \frac{1}{2}  \omega_{i}^{2} \Bigl(\vect{q}_{i} - \frac{c_{i}}{ \omega_{i}^{2}} \vect{Q} \Bigr )^{2} + \dot{\gamma} q_{iy} p_{ix}\hat{\vect{x}}\hat{\vect{y}}\Bigr],
\end{equation}
respectively.

We compute the time-correlation function of the noise for both of the cases discussed above. 
As the noise function has two different expressions for every component, the integration is carried out on $q_x(0),p_x(0)$ and $q_y(0),p_y(0)$ separately. For the equilibrium initial conditions, the Boltzmann weight does not mix $\vect{q}(0)$'s and $\vect{p}(0)$'s  $x$ and $y$ components as there is no the shear term that couples them, so they can be regarded as independent random variables. That is, given two functions depending on one component only  $f(q_x(0),p_x(0)), g(q_y(0),p_y(0))$ the  average of their product is equal to the product of the single averages:

\begin{multline}
 \langle f(q_x(0),p_x(0)) g(q_y(0),p_y(0)) \rangle \\= \langle f(q_x(0),p_x(0)) \rangle \langle g(q_y(0),p_y(0))\rangle.
\end{multline}
 Consequently, every integral of the form
\begin{multline}
\int f(q_x(0),p_x(0)) g(q_y(0),p_y(0)) e^{-\Ham_B/k_B T}\\ \times dq_x(0) dp_x(0) dq_y(0) dp_y(0)  
\end{multline}
\noindent
can be evaluated as a product of integrals carried out on single Cartesian components:
\begin{multline}
\int \Bigl(f(q_x(0),p_x(0)) g(q_y(0),p_y(0))\Bigr) e^{-\Ham_B/k_B T}dq_x(0) \\
\times dp_x(0) dq_y(0) dp_y(0) \\= \int f(q_x(0),p_x(0))e^{-\Ham_B/k_B T} dq_x(0) dp_x(0) \\\int g(q_y(0),p_y(0))e^{-\Ham_B/k_B T}dq_y(0) dp_y(0).
\end{multline}

In the case of nonequilibrium initial conditions, the positions and momenta of the heat bath's oscillators are correlated random variables as the shear term couples $p_x(0)$ and $q_y(0)$ so in that case we have integrals of the form:

\begin{multline}
 \langle f(q_x(0),p_x(0)) g(q_y(0),p_y(0)) \rangle = \\
\int f(q_x(0),p_x(0)) g(q_y(0),p_y(0)) e^{-\Ham_B/k_B T}\\ \times dq_x(0) dp_x(0) dq_y(0) dp_y(0).
\end{multline}

\section{Results}
We begin by solving the heat bath's equations of motion. They form a system of coupled ordinary differential equations. In order to be solved, the system needs to be diagonalized first.

By looking at the eigenvalues, in Eq. \eqref{eigenvalues}, one of them is always imaginary while the other one can be either real or imaginary  depending on the shear rate's intensity compared to the harmonic oscillators' frequencies. If the shear rate is lower than the oscillators' frequency, the system's solution will be a linear combination of trigonometric functions, that is, the heat bath's degrees of freedom will oscillate with an oscillation frequency modified by the shear-rate. However, if the shear-rate is higher than the oscillators' frequency, the shear will affect the overall behaviour.

Consequently, we present the results we obtained by distinguishing two different cases or limits: $\dot{\gamma} < \omega_i   \quad   \forall i$ and $\dot{\gamma} > \omega_i \quad \forall i$. 
As in natural units the vibrational frequency equals the vibrational temperature, it is also possible to relate the shear rate's intensity to the average vibrational temperature of the heat-bath's molecules. 

\begin{figure}[h!]
    \includegraphics[width=0.45\textwidth]{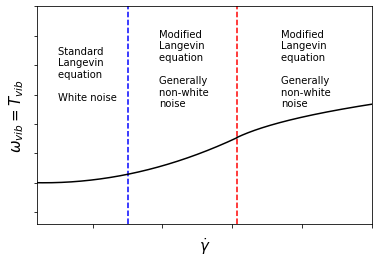}
    \caption{Qualitative phase diagram of the obtained results for the different forms of the Langevin equation with shear. The solid line indicates the schematic dependence of the vibrational temperature (frequency) of the thermal bath as a function of the applied shear rate across three different regimes discussed in the text: to the left, $\dot{\gamma}\ll\omega_{i}$, in the middle $\dot{\gamma}<\omega_{i}$, and to the right $\dot{\gamma}>\omega_{i}$. }
    \label{fig:diagram}
\end{figure}

This allows one to visualize the obtained results drawing a qualitative phase diagram presented in Fig. \ref{fig:diagram}. We first present the results graphically and then we explain them in detail in the next sections.

\pagebreak

\subsection{Generalized Langevin equation with shear}
\subsubsection{Close to equilibrium: $\dot{\gamma} \ll \omega_i$  and $\dot{\gamma}<\omega_i$}

If $\gamma \ll \omega_i$ all the terms like $\dot{\gamma} +\omega_i$ or $\dot{\gamma} -\omega_i$ can be approximated to $\simeq \omega_i$.  In this case the Langevin equation reduces to

\begin{equation} \label{lang_ssmall}
    \begin{split}
  \dot{\vect{P}} = -V'[\vect{Q}]+\vect{F}(t) -\int_0^t K(t-t')\dot{\vect{Q}}(t')dt' -\dot{\gamma}P_x \hat{\vect{y}}   
 \end{split}
\end{equation}

\noindent where 
\begin{equation}
    K(t-t') \equiv \nu(t-t') =\sum_i \frac{c_i^2}{\omega_i^2} \cos(\omega_i(t-t')).\label{memoryfunction_1}
\end{equation}

\noindent This equation is very similar to Eq. \eqref{lang_simultions_1} widely used in the literature \cite{zwanzig2001nonequilibrium}.
In particular, the memory kernel is exactly the same as the one derived by Zwanzig for the GLE without shear \cite{zwanzig2001nonequilibrium}.

There are, however, some remarkable differences which will be analyzed in the following sections. As a sanity check, it is important to notice that if $\dot{\gamma}$ is set equal to zero, then the equation for the Brownian particle in a solvent with no external drag forces is recovered. In particular, the memory function $\nu(t-t')$ is precisely the same, as all the dependence on $\dot{\gamma}$ is encoded in the drag term $\dot{\gamma} P_x \hat{\vect{y}}$. However, the drag term is different because the friction does not appear as a factor. Also, this drag force is oriented along the $\hat{\vect{y}}$-axis direction and not along the $\hat{\vect{x}}$-axis direction. 

If the shear rate is less than the heat bath's frequencies but not entirely negligible, we expect the fluid's particles to oscillate around their equilibrium positions with a frequency modified by the shear rate. As we are in the limit of small shear rate, we can perform a Taylor expansion around $\dot{\gamma} = 0 $. If only terms up to first order are kept, the Langevin equation reads as

\begin{equation}   \label{lang_small_1}
    \begin{split}
         \dot{\vect{P}} &= -V'[\vect{Q}] - \int_0^t \nu(t-t') \dot{\vect{Q}}(t') dt'\\
         &- \dot{\gamma} \int_0^t \vect{K}_1(t-t') \dot{\vect{Q}}(t')dt' 
         - \dot{\gamma} \int_0^t \vect{K}_2(t-t') \dot{\vect{Q}}(t')dt'\\
        & + \vect{F}(t) - \dot{\gamma} P_x(t) \hat{\vect{y}} + o(\dot{\gamma})
   \end{split}
\end{equation}


\noindent where $\vect{K}_1(t-t')$ and $\vect{K}_2(t-t')$ are friction matrices, which have the following form

\begin{widetext}

\begin{equation}
\vect{K}_1(t-t') = \begin{pmatrix} 0 & \sum_i \frac{c_i^2}{2\omega_i^2} \cos\left[ \omega_i (t-t')\right] (t-t') \\ - \sum_i \frac{c_i^2}{2\omega_i^2} \cos\left[\omega_i  (t-t')\right] (t-t') & 0 
\end{pmatrix}
\end{equation}

 \begin{equation}
     \vect{K}_2 (t-t')= \begin{pmatrix}
      0 & - \sum_i \frac{c_i^2}{2\omega_i^3} \sin\left[ \omega_i  (t-t')\right]  \\ \sum_i \frac{c_i^2}{2\omega_i^3} \sin\left[ \omega_i  (t-t')\right] & 0 
     \end{pmatrix}
 \end{equation}
 
\end{widetext}

\noindent and $\nu(t-t')$ is the memory function in the absence of shear, introduced above.


\noindent At first order, the shear rate does not affect the oscillation frequencies but introduces some additional friction terms which are, in general, non-Markovian as they cannot be easily reduced to be proportional to a Dirac's delta function.

If also second order terms are kept, the GLE becomes
\begin{widetext}

\begin{equation}  \label{lang_small}
    \begin{split}
         \dot{\vect{P}} &= -V'[\vect{Q}] - \int_0^t K(t-t') \dot{\vect{Q}}(t') dt' - \dot{\gamma} \int_0^t \vect{K}_1(t-t') \dot{\vect{Q}}(t')dt' - \dot{\gamma} \int_0^t \vect{K}_2(t-t') \dot{\vect{Q}}(t')dt' 
 \\ &-\dot{\gamma}^2 \int_0^t  \vect{K}_3(t-t') \dot{\vect{Q}}(t')dt'  
 + \vect{F}(t) - \dot{\gamma} P_x(t) \hat{\vect{y}} + o(\dot{\gamma}^2)
   \end{split}
\end{equation}

\end{widetext}
\noindent where the friction matrices read as:

\begin{widetext}

\begin{equation}
\vect{K}_1(t-t') = \begin{pmatrix} 0 & \sum_i \frac{c_i^2}{2\omega_i^2} \cos\left[ \Bigl(\omega_i - \frac{\dot{\gamma}^2}{8\omega_i} \Bigr) (t-t')\right] (t-t') \\ - \sum_i \frac{c_i^2}{2\omega_i^2} \cos\left[ \Bigl(\omega_i - \frac{\dot{\gamma}^2}{8\omega_i} \Bigr) (t-t')\right] (t-t') & 0 
\end{pmatrix}
\end{equation}

 \begin{equation}
     \vect{K}_2 (t-t')= \begin{pmatrix}
      0 & - \sum_i \frac{c_i^2}{2\omega_i^3} \sin\left[\Bigl( \omega_i - \frac{\dot{\gamma}^2}{8\omega_i} \Bigr) (t-t') \right]  \\ \sum_i \frac{c_i^2}{2\omega_i^3} \sin\left[\Bigl( \omega_i - \frac{\dot{\gamma}^2}{8\omega_i} \Bigr) (t-t') \right] & 0 
     \end{pmatrix}
 \end{equation}

 \begin{equation}
     \vect{K}_3(t-t') = 
     \begin{pmatrix}
         0 & 0 \\ 0 & \sum_i \frac{c_i^2}{\omega_i^4} \cos\left[\Bigl( \omega_i - \frac{\dot{\gamma}^2}{8\omega_i} \Bigr) (t-t') \right] + \frac{c_i^2}{2\omega_i^3}(t-t') \sin\left[\Bigl( \omega_i - \frac{\dot{\gamma}^2}{8\omega_i} \Bigr) (t-t') \right]
     \end{pmatrix}
 \end{equation}
 
\end{widetext} 
\noindent and 

 \begin{equation} 
 \begin{split}
     &K(t-t') =\\
     &\sum_i \frac{c_i^2}{\omega_i^2} \cos\Bigl(\Bigl( \omega_i - \frac{\dot{\gamma}^2}{8\omega_i}\Bigr) (t-t') \Bigr) \cos\Bigl(\frac{\dot{\gamma}}{2}(t-t') \Bigr) 
     + o(\dot{\gamma}^2).\label{memory_function}
     \end{split}
 \end{equation}
 
\balance

By looking both at the friction matrices and at the memory function, we see that the shear rate  introduces a second order correction in the heat bath's oscillation frequencies. Moreover, the memory function Eq.~\eqref{memory_function} differs from the first-order one Eq.~\eqref{memoryfunction_1} both for the oscillation frequency and for the new expression as, in this case, there's a product with a function depending on $\dot{\gamma}$. As

\begin{equation}
    \cos\Bigl(\frac{\dot{\gamma}}{2}(t-t')\Bigr) \simeq 1 - \frac{1}{2} \Bigl(\frac{\dot{\gamma}}{2}(t-t')\Bigr)^2  + o(\dot{\gamma}^2)
\end{equation}

\noindent the memory function $K(t-t')$ can also be rewritten as

\begin{equation}
\begin{split}
  &K(t-t') = \\ 
  &\sum_i \frac{c_i^2}{\omega_i^2}\Biggl[ \Bigl(1- \frac{\dot{\gamma}^2}{8} \Bigr)(t-t')^2\cos\Bigl(\Bigl( \omega_i - \frac{\dot{\gamma}^2}{8\omega_i}\Bigr) (t-t') \Bigr) \Biggr]. 
   \end{split}
\end{equation}

If all the terms proportional to powers of  $\dot{\gamma}$ are grouped in the same friction matrix, Eq. \eqref{lang_small} can be also written as a more compact expression as shown in the Appendix A. However, the former expression makes the interpretation of the corresponding FDT clearer, as we shall see in Sec. III.B, since making the memory function explicit allows us to give a more physical interpretation to the FDT. 


As in the previous case, by setting $\dot{\gamma} = 0$ the Langevin equation for the Brownian particle with no shear is recovered because all the additional terms proportional to $ \dot{\gamma}$ vanish and the heat bath's oscillation frequencies become identically equal to $\omega_i$.

\subsubsection{Far from equilibrium: $\dot{\gamma} \gg \omega_i \quad \forall i$}

We present the GLE obtained by considering the same system far from equilibrium. That is, we assume the external shear perturbation to be higher than the thermal agitation. If we take $\dot{\gamma} \gg \omega_i \quad \forall i = 1,...,N$ then  $\dot{\gamma} \pm \omega_i \simeq \dot{\gamma}$ and Eq.\eqref{lang_gen} becomes


\begin{equation}
 \begin{split}
 \dot{\vect{P}} &= -V'[\vect{Q}] - \int_0^t \vect{\overline{K}}_1(t-t') \dot{\vect{Q}}(t') dt' \\
 &-\int_0^t \vect{\overline{K}}_2(t-t') \dot{\vect{Q}}(t')dt' - \dot{\gamma} P_x(t) \hat{\vect{y}} 
  \end{split}
\end{equation}

 \noindent where the friction matrices now read as 
 
\begin{widetext}

 \begin{equation}
    \vect{\overline{K}_1} (t-t')= \begin{pmatrix}
      0 &  \sum_i \frac{c_i^2}{2\omega_i\sqrt{\dot{\gamma}\omega_i}} \Bigl[\sin( \sqrt{\dot{\gamma}\omega_i}(t-t') ) - \sinh( \sqrt{\dot{\gamma}\omega_i}(t-t') )\Bigr]\\ \sum_i \frac{c_i^2}{2\omega_i\sqrt{\dot{\gamma}\omega_i}} \Bigl[-\sin( \sqrt{\dot{\gamma}\omega_i}(t-t') )+\sinh( \sqrt{\dot{\gamma}\omega_i}(t-t') )\Bigr]  & 0 \end{pmatrix}
       \end{equation}
      
      \begin{equation}
     \vect{\overline{K}_2} (t-t')= \begin{pmatrix}
    \sum_i  \frac{c_i^2}{\omega_i^2}\Bigl[ -\cosh(\sqrt{\dot{\gamma}\omega_i}(t-t') ) + \cos(\sqrt{\dot{\gamma}\omega_i}(t-t') )\Bigr]& 0 \\ 0 &  \sum_i \frac{c_i^2}{\omega_i\dot{\gamma} }\Bigl[-\cosh(\sqrt{\dot{\gamma}\omega_i}(t-t') ) + \cos(\sqrt{\dot{\gamma}\omega_i}(t-t') )\Bigr] \end{pmatrix}.
     \end{equation}

\end{widetext}

\noindent As in the previous case, we find a GLE 
with different memory functions. In this case, the memory functions reflect the effect of the external driving force: on the one hand there's an exponential growth (due to the $\cosh$) and on the other hand an oscillation around the equilibrium positions with a modified frequency. 
As before, it is possible to write the Langevin equation by grouping all the friction matrices like

\begin{widetext}

\begin{equation}
 \begin{split}
 \dot{\vect{P}} &= -V'[\vect{Q}] - \int_0^t \vect{\overline{K}}(t-t') \dot{\vect{Q}}(t') dt' - \dot{\gamma} P_x(t) \hat{\vect{y}} 
  \end{split}
\end{equation}

where
\begin{equation}
    \vect{\overline{K}}_1 (t-t')= \begin{pmatrix}
       \sum_i  \frac{c_i^2}{\omega_i^2}\Bigl[ \cosh(\sqrt{\dot{\gamma}\omega_i}(t-t') ) + \cos(\sqrt{\dot{\gamma}\omega_i}(t-t') )\Bigr]
      &  \sum_i \frac{c_i^2}{2\omega_i\sqrt{\dot{\gamma}\omega_i}} \Bigl[\sin( \sqrt{\dot{\gamma}\omega_i}(t-t') ) - \sinh( \sqrt{\dot{\gamma}\omega_i}(t-t') )\Bigr] \\ \sum_i \frac{c_i^2}{2\omega_i\sqrt{\dot{\gamma}\omega_i}} \Bigl[-\sin( \sqrt{\dot{\gamma}\omega_i}(t-t') ) + \sinh( \sqrt{\dot{\gamma}\omega_i}(t-t') )\Bigr]  &  \sum_i \frac{c_i^2}{2\dot{\gamma}\omega_i}\Bigl[ \cosh(\sqrt{\dot{\gamma}\omega_i}(t-t') ) + \cos(\sqrt{\dot{\gamma}\omega_i}(t-t') )\Bigr] \end{pmatrix}.
       \end{equation}

\end{widetext}

\noindent Also in this case we see how the external shear force affects the dynamics at every level of description, since the new oscillation frequency depends on $\dot{\gamma}$ as $\sqrt{\dot{\gamma}}$. 

\subsection{Fluctuation-dissipation theorems}
Starting from the noise's expression Eq.~\eqref{noise_x} and Eq.~\eqref{noise_y}, we compute a FDT both for equilibrium and nonequilibrium heat bath's initial conditions. In doing so, different cases are analyzed.
We still distinguish between high and low shear rate, but also between equilibrium and nonequilibrium initial conditions. That is, we compute a FDT for small shear rates both for a system already perturbed at $t=0$ and for an unperturbed one in equilibrium at $t=0$, respectively. We then present some physical considerations about the behavior for high shear rates and obtain an analytical result in the long time limit. 

\subsubsection{Equilibrium initial conditions}
If the system is assumed to be in equilibrium at $t=0$ and the external drag force acts from $t>0$ on, when computing $\langle \vect{F}(t) \vect{F}(t') \rangle$ the Boltzmann weight is given by the heat bath's Hamiltonian with no shear Eq.~\eqref{ham_noshear}. If only first order terms are kept, we find

\begin{equation} \label{FDT_eq_1}
 \langle\vect{F}(t) \vect{F}(t') \rangle = k_B T \nu(t-t') + o(\dot{\gamma})
\end{equation}

 \noindent where the memory function $\nu(t-t')$ is given by Eq.~\eqref{memoryfunction_1}. This is precisely the same fluctuation-dissipation relation for a Brownian particle in a quiescent system (i.e. in the absence of external shear) derived e.g. in \citep{zwanzig2001nonequilibrium}. This is also reassuring as a sanity check. Therefore, all the considerations already done about this expression are still valid. If also second order terms are considered, then the fluctuation-dissipation theorem (cfr. Appendix B.1 for the full details) becomes

 \begin{multline} \label{FDT_eq}
 \langle\vect{F}(t) \vect{F}(t') \rangle = k_B T \nu(t-t') + \dot{\gamma}^2 k_B T \vect{H}(t,t',\omega_i)\\+ \dot{\gamma}^2 k_B T \nu(t-t') \hat{\vect{y}}+o(\dot{\gamma}^2)
 \end{multline}
 
 \noindent where 
 
 \begin{multline}
     \vect{H}(t,t',\omega_i) = \sum_i \frac{c_i^2}{8\omega_i^3}\Bigl[\sin(\omega_i(t+t'))(t-t')\\-\sin(\omega_i(t-t'))(t+t')\Bigr].
 \end{multline}
 
This expression Eq.\eqref{FDT_eq} is made of a term proportional to the memory function $\nu(t-t')$ Eq.\eqref{memory_function}  plus two second-order corrections. One term is proportional to a function $\vect{H}(t,t',\omega_i)$ depending both on $t-t'$ and on $t+t'$, that is, not only on time differences but also on the entire time evolution of the system. The other term is proportional to $\dot{\gamma}^2$ and to the friction function $\nu(t-t')$, so it stands for the standard energy dissipation as it occurs in sheared fluids \citep{landau1959fluid}.  

 Therefore, up to linear order in $\dot{\gamma}$, this result is consistent with the hypotheses that are usually assumed in numerical simulations \citep{ikeda2012unified}. However, at second order in $\dot{\gamma}$ this is no longer true. The dissipative term quadratic in $\dot{\gamma}$ is  consistent with the physical picture of the system, as there is a term representing transverse energy dissipation as one expects for a Couette flow as different fluid layers move parallel with a velocity directed along the $x$ axis so that energy dissipation occurs along the $y$ axis.

\subsubsection{Nonequilibrium initial conditions}

We now assume the system to be already perturbed at $t=0$. Heat-bath's initial conditions are then Boltzmann-distributed with a weight given by Eq.~\eqref{hamiltonian}. At first order we find

\begin{equation}
    \langle\vect{F}(t) \vect{F}(t')\rangle = k_B T \nu(t-t')  + \dot{\gamma}\Tilde{\vect{G}}(t,t',\omega_i) + o(\dot{\gamma})
\end{equation}

\noindent where

\begin{widetext}

\begin{equation} 
\Tilde{\vect{G}}(t,t',\omega_i) = \begin{pmatrix} G_x(t,t',\omega_i) \\ G_y(t,t',\omega_i) 
     \end{pmatrix} \\
      = \begin{pmatrix} \sum_i\Bigl[ \frac{c_i^2}{2\omega_i^2}\sin(\omega_i(t+t')) (t+t')+\frac{c_i^2}{2\omega_i^2} (t-t') \cos(\omega_i (t-t'))\Bigr]  \\ \sum_i \Bigl[\frac{c_i^2}{2\omega_i^3}\cos(\omega_i(t+t'))+\frac{c_i^2}{2\omega_i^3}   \cos(\omega_i(t-t'))\Bigr]
     \end{pmatrix}
\end{equation}

\end{widetext}

\noindent The main difference respect to Eq.~\eqref{FDT_eq_1} is that here there are some corrections at first order in $\dot{\gamma}$ already. Although these corrections break parity symmetry, it is important to remark that parity is broken already in the initial state, as it is indicated by the Hamiltonian Eq.~\eqref{hamiltonian}. As in Eq.~\eqref{FDT_eq_1}, one of the additional terms is proportional to a function $\Tilde{\vect{G}}(t,t',\omega_i)$ that depends on the entire time-evolution of the system. If also second order terms in $\dot{\gamma}$ are kept, the FDT reads as (cfr. Appendix B.2 for the full details)

\begin{multline}\label{FDT_noneq}
    \langle\vect{F}(t) \vect{F}(t')\rangle = k_B T \nu(t-t')  + \dot{\gamma}\Tilde{\vect{G}}(t,t',\omega_i) \\+ \dot{\gamma}^2 k_B T \Tilde{\vect{H}}(t,t',\omega_i) + \dot{\gamma}^2 k_B T \nu(t-t') \hat{\vect{y}} + o(\dot{\gamma}^2)
\end{multline}

\noindent where 

\begin{widetext}

\begin{equation}
\begin{split}
&\Tilde{\vect{H}}(t,t',\omega_i) = \begin{pmatrix} \Tilde{H}_x(t,t',\omega_i) \\ \Tilde{H}_y(t,t',\omega_i) \end{pmatrix} \\
&= \begin{pmatrix} \sum_i \Bigl[\frac{c_i^2}{8\omega_i^4} \cos\Bigl(\Bigl(\omega_i - \frac{\dot{\gamma}^2}{8\omega_i^2}\Bigr)(t-t')) \cos \Bigl(\frac{\dot{\gamma}}{2}(t+t')\Bigr) + \cos \Bigl(\Bigl(\omega_i - \frac{\dot{\gamma}^2}{8\omega_i^2}\Bigr)(t+t')\Bigr)\cos \Bigl(\frac{\dot{\gamma}}{2}(t-t')\Bigr) \Bigr] \\ \sum_i\frac{3 c_i^2}{2\omega_i^4} \cos((\omega_i - \frac{\dot{\gamma}^2}{8\omega_i})(t+t'))\cos(\frac{\dot{\gamma}}{2}(t+t'))    \end{pmatrix} 
\end{split}
\end{equation}

\end{widetext}

\noindent
As before, the function $\Tilde{\vect{H}}(t,t',\omega_i)$ depends also on the time-evolution of the system. It is important to notice that also here there is a term that represents viscous fluid energy dissipation \cite{landau1959fluid}.  

\subsection{Far from equilibrium ($\dot{\gamma}\gg\omega_i$) and turbulence} 
We also calculated a FDT in the case of strong external shear perturbation. In this case, the oscillation frequencies are replaced by $\sqrt{\dot{\gamma}\omega_i}$. We focused on the corresponding long-time limit. The final expression is lengthy and it features a linear combination of products between hyperbolic functions and trigonometric functions. For this reason, it is not possible to determine, analytically, whether the $t \rightarrow \infty$ limit converges or not. For high shear rates, the flow is expected to become turbulent, so the model used here is expected to fail. Therefore, it is necessary to determine when the transition from laminar to turbulent flow happens and to relate this to the shear rate's intensity. This can be done by writing the Reynolds number as a function of the shear rate. It is known, in fact, that a flow is laminar as long as the Reynolds number is lower than a threshold, which depends on the characteristics of the device, and  becomes turbulent above the threshold. The Reynolds number is defined as a ratio of the inertial forces to the viscous forces

\begin{equation}
Re = \frac{\rho u d}{\mu}
\end{equation}

\noindent where $\rho$ is the fluid's density, $\mu$ is the fluid's kinematic viscosity, $d$ is the diameter of the considered fluid's section and $u$ is the flow's velocity. Since for e.g. a Couette flow the flow velocity is known and proportional to the shear rate, the Reynolds number can also be written as

\begin{equation} \label{reynolds}
Re(\dot{\gamma} )=  \frac{\rho \dot{\gamma} y d}{\mu}.
\end{equation}
 
As a laminar flow occurs when $Re < 2300$, this threshold gives an upper bound on the shear rate's intensity up to which the above model is applicable. It depends also on the particular fluid the model is applied to, as in Eq.\eqref{reynolds} also some parameters depend crucially on the fluid's physico-chemical characteristics.


\section{Discussion}

\subsection{Generalized Langevin equation with shear}
The first task we addressed was the derivation of a generalized Langevin equation (GLE) with shear flow from first principles. We found an equation that is, in general, different from the one used in literature. However, we notice that in the limit of extremely low shear-rate it has a form that is very similar to the latter one, except for the drag term. This is because the drag term is directed along the $\hat{\vect{y}}$ axis, and is not multiplied by a friction coefficient. The friction coefficient is implicit in $P_x(t)$'s analytic expression as it can be obtained by integrating the same equations that give the memory functions. Instead, the fact that the drag term is directed along $\hat{\vect{y}}$, and not along $\hat{\vect{x}}$, is a feature of the DOLLS dynamics. It would be possible to obtain the same equation used in the literature by rotating the reference system, but it would affect the physical picture of the other results, such as the transverse energy dissipation. 

We performed a perturbative expansion in series of $\dot{\gamma}$ centered in $\dot{\gamma}=0$ as it points out  the effect of the external shear force by treating it as a perturbation. It shows that the additional memory functions appear as perturbative corrections to the already known equation, as they are proportional to power of $\dot{\gamma}$. In the GLE Eq.\eqref{lang_small} there is also a term (not shown) that apparently breaks translational invariance as it is proportional to the tagged particle's position $\vect{Q}(t)$. It appears as a second order correction so, in first approximation, it can be ignored.

It is remarkable that if $\dot{\gamma}$ is set equal to zero, the same Langevin equation for a Brownian particle with no external perturbation is recovered \citep{zwanzig2001nonequilibrium}. We also investigated how the GLE changes when the external shear force is higher compared to the bath's thermal agitation. We found an equation which has the same form as the one with small shear Eq.\eqref{lang_small} but with different friction matrices. In fact, as we assumed $\dot{\gamma} > \omega_i $ $\forall i$, one of the eigenvalues is real so the memory functions are not all trigonometric functions but there are also hyperbolic functions that express the exponential expansion of the system determined by the high external driving force.  
We can conclude that we derived a GLE which can be used either close to and far from equilibrium and that, in the special case of a system very close to equilibrium, is similar to the known Langevin equation with shear used in the literature. How to more closely derive the latter is shown in the following subsection.

\subsubsection*{Relation to the Langevin equation with shear used in numerical simulation studies}
In the previous sections it was stated that the results presented above were based on a system where both the colloidal particle and the heat-bath's degrees of freedom were subjected to the external force field. This led to a generalized Langevin equation which is, in general, different from the one that can be found in literature. It is therefore relevant to investigate whether the same equation can be obtained assuming different hypothesis. If the external force field is assumed to act only on the heat bath's degrees of freedom and not directly on the tagged particle, different equations of motion are found because the tagged particle's Hamiltonian won't have the additional DOLLS term. In this case, the dynamics of the tagged particle is simply the Caldeira-Leggett one:

\begin{equation} 
\begin{split}
&\dot{\vect{Q}} = \frac{\vect{P}}{M} \\
&\dot{\vect{P}} =-V'[\vect{Q}] +\sum_i c_i\Bigl(\vect{q}_{i}-\frac{c_i \vect{ Q}}{\omega_i^2}\Bigr).
\end{split}
\end{equation}

By proceeding in the same way as before and plugging heat bath's solutions Eq.\eqref{sol_bath_x} and Eq.\eqref{sol_bath_y} into these equations of motion, a different Langevin equation with shear is found. In this case, indeed, the somewhat unphysical drag term $\dot{\gamma} P_x \hat{\vect{y}}$ is now absent, which is clear since the tagged particle does not experience any direct perturbation due to the externally applied field. 

Furthermore, the previous Langevin equation was obtained by integrating by parts all the integral terms appearing in Eq.\eqref{sol_bath_x} and Eq.\eqref{sol_bath_y}. However, if the integration of the heat bath's $x$ component Eq.\eqref{sol_bath_x} is carried over the $Q_x$ integrals only while all the integral terms appearing in the $y$ component Eq.\eqref{sol_bath_y} are integrated as before, the following Langevin equation with shear is found:

\begin{equation}\label{correct}
     \begin{split}
  \dot{\vect{P}} =& -V'[\vect{Q}]+\vect{F}(t) -\int_0^t \nu(t-t') \dot{\vect{Q}}(t')dt' \\&+\dot{\gamma} \int_0^t \eta(t-t') Q_y(t') dt'    \hat{\vect{x}} \\& + \dot{\gamma} \int_0^t \mu (t-t') \dot{Q}_x(t')dt'   \hat{\vect{y}}
 \end{split}
\end{equation}

\noindent where $\eta(t-t')$ and $\mu (t-t')$ are new memory functions:

\begin{equation}
    \eta (t-t') = \sum_i \Biggl[\frac{c_i^2}{\omega_i} (t-t') \sin(\omega_i(t-t'))\Biggr] 
    \end{equation}
    
    \begin{equation}
        \begin{split}
         \mu(t-t') =& \sum_i \Biggl[ \frac{c_i^2}{2\omega_i^2}(t-t') \cos(\omega_i(t-t'))\\& - \frac{c_i^2}{4\omega_i^3}\sin(\omega_i(t-t')) \Biggr].\label{spurious}
    \end{split}
\end{equation}

This equation has the right drag term, which is  proportional to the $y$ component of the tagged particle's position and is directed along the longitudinal $x$ axis, but it still has an additional term (the last one on the r.h.s. of Eq. \eqref{correct})  that does not appear in Eq. \eqref{lang_simultions_1}. However, this term becomes negligible if the high frequency limit is taken, as the terms in Eq. \eqref{spurious} are proportional to negative higher-order ($>1$) powers of the oscillation frequencies $\omega_{i}$. 
In this limit, we thus get:
\begin{equation}\label{correct}
     \begin{split}
  \dot{\vect{P}} =& -V'[\vect{Q}]+\vect{F}(t) -\int_0^t \nu(t-t') \dot{\vect{Q}}(t')dt' \\&+\dot{\gamma} \Biggl[\int_0^t dt' \eta(t-t') Q_y(t') dt' \Biggr]   \hat{\vect{x}}.
 \end{split}
\end{equation}

Consequently, this equation recovers the generic form of the Langevin equation that can be found in the literature under the above stated assumptions. The main difference is that, in this case, there is a strongly non-Markovian behavior as the drag term (proportional to $\dot{\gamma}$) cannot be easily reduced to be proportional to a Dirac delta function. In other words, we found a generalized Langevin equation with shear flow which is the non-Markovian equivalent of that used in the literature \cite{berthier2002shearing}.

This Langevin equation
 has a noise function that is slightly different from the one in the previous Langevin equation (Eq.\eqref{lang_small_1}), but this does not affect the FDT. In fact, the differences are proportional to the tagged particle's initial position and, as it was stated before, the dependence on them has been suppressed because, with no loss of generality, the tagged particle can be assumed to be initially at the origin of the reference framework. For this reason, the FDT reads as

 \begin{multline} 
 \langle\vect{F}(t) \vect{F}(t') \rangle = k_B T \nu(t-t') + \dot{\gamma}^2 k_B T \vect{H}(t,t',\omega_i)\\+ \dot{\gamma}^2 k_B T \nu(t-t') \hat{\vect{y}} + o(\dot{\gamma}^2)
 \end{multline}
in the case of equilibrium initial conditions and 
 \begin{multline}
    \langle\vect{F}(t) \vect{F}(t')\rangle = k_B T \nu(t-t')  + \dot{\gamma}\Tilde{\vect{G}}(t,t',\omega_i) \\+ \dot{\gamma}^2 k_B T \Tilde{\vect{H}}(t,t',\omega_i) + \dot{\gamma}^2 k_B T \nu(t-t') \hat{\vect{y}} + o(\dot{\gamma}^2)
\end{multline}
with nonequilibrium initial conditions (see Eq.\eqref{FDT_eq} and Eq.\eqref{FDT_noneq} for more details).
 
 There is some arbitrariness in the procedure described above, as the terms to be integrated out in the dynamics have been chosen in an ad hoc way. This particular choice implies ignoring the dynamics of microscopic degrees of freedom along the $y$-axis direction (while not along the $x$-axis direction), which could be tentatively justified based on the physics of the system since the external shearing force is directed along the $x$ axis. This axis thus represents the ``special'' direction along which the dynamics is being tracked and cannot be integrated out together with the other degrees of freedom.

\subsection{Fluctuation-dissipation theorems}
From the explicit noise expression we computed the associated fluctuation-dissipation theorem (FDT). The first issue that has been encountered was the choice of the initial distribution. In fact, depending on the form of the Boltzmann weight, different physical situations are described and because of this sort of arbitrariness we decided to analyze two different paradigmatic cases. 

It is important to point out that, depending on the choice of initial distribution, different results are found, which is expected for out-of-equilibrium systems. In the following discussion we will assume independence from the tagged particle's initial conditions. In fact, it can be assumed, with no loss of generality, that the tagged particle is at the origin of the coordinate frame at $t=0$. 

At first we analyzed a situation of ``start-up'' shear, that is, we assumed the external driving force to not act at $t=0$ but only at $t>0$. As for the generalized Langevin equation with shear, under these assumptions we found a FDT that recovers the classical one with no shear for low shear-rates. By performing a perturbative expansion around $\dot{\gamma}=0$, in the same way as we did for the GLE, and assuming a marginally higher shear-rate, we found a different FDT with two additional terms. They are both second-order corrections as they are proportional to $\dot{\gamma}^2$. The first term contains a function $\vect{H}(t,t',\omega_i)$ that depends on the entire time evolution of the system, while the second correction term is proportional also to a memory function and represents the ordinary energy-dissipation in a viscous fluid \citep{landau1959fluid}. Moreover, the memory function differs from the one that appears in \citep{zwanzig2001nonequilibrium} for the zero-shear case. As Eq. \eqref{FDT_eq} shows a violation of the ordinary FDT (as shown already with different methods in Refs. \cite{Szamel1,Szamel2,berthier2002nonequilibrium,Rubi,Leporini}), one could ask whether it represents a fluctuation-dissipation relation or not, but as there is a term compatible with the energy dissipation in fluids, it is meaningful to say that Eq. \eqref{FDT_eq} still relates the noise's fluctuations with the energy dissipation. Moreover, if $\dot{\gamma}$ is set equal to zero, the ordinary zero-shear FDT is correctly recovered. 
When assuming the system not to be already perturbed at $t=0$, a different FDT is found. As shown by Eq.\eqref{FDT_noneq}, there are some corrections at first order in $\dot{\gamma}$ already, which thus break parity and time-reversal symmetry (which already occurs in the initial nonequilibrium state). The other first-order correction is proportional to a vector function that depends on the entire time evolution of the system. On the other hand, second order corrections are similar to those in Eq. \eqref{FDT_eq} except for $\vect{H}$ that has a different expression. Consequently, if nonequilibrium conditions are assumed, the corresponding fluctuation-dissipation relation violates the usual FDT already at first order in $\dot{\gamma}$ (cfr., with different methods, Refs. \cite{Szamel1,Szamel2,berthier2002nonequilibrium,Rubi,Leporini}).

\subsection{Markovianity}
The Langevin dynamics with shear is usually assumed, throughout the literature, to be a Markov process. If $\dot{\gamma} \ll \omega_i$, second order corrections appearing in the FDT expression Eq.\eqref{FDT_eq} can be ignored and therefore the ordinary FDT for the Brownian particle \citep{zwanzig2001nonequilibrium} is recovered. However, already in the limit  $\dot{\gamma} < \omega_i$, the FDT contains some corrections proportional to functions depending on the entire time-evolution of the system. This indicates that, in general, this process is not Markovian since, by definition, in a Markov process the dynamics at a fixed time $t$ is determined only by the dynamics of the system in a previous time instant and not by the entire time-evolution. If the Taylor expansion is truncated at first order, we recover the same FDT as in Zwanzig's derviation\citep{zwanzig2001nonequilibrium}. Consequently, by taking the continuum limit

\begin{equation}
    \sum_i \rightarrow \int d\omega g(\omega).
\end{equation}

\noindent with 

\begin{equation}
    g(\omega) \simeq \omega^2
\end{equation}
as for bosonic particles, we find
\begin{equation}
\langle \vect{F}(t) \vect{F}(t') \rangle \simeq k_B T \delta(t-t')
\end{equation}

so that the process is Markovian. If also second-order terms in $\dot{\gamma}$ are kept, it is no longer Markovian because of the function $\vect{H}(t,t',\omega_i)$. Since   $\vect{H}(t,t',\omega_i)$ is proportional to $\frac{1}{\omega^3}$, the integral 

\begin{equation}
\int_0^\infty d\omega g(\omega)\vect{H}(t,t',\omega)
\end{equation}

\noindent diverges close to $\omega = 0$ unless the frequencies distribution (i.e. the density of states) $g(\omega)$ has a different dependence on $\omega$.

When nonequilibrium initial conditions are chosen, the process is strongly non-Markovian at first order in $\dot{\gamma}$ already. As in the previous case, when the continuum limit is performed, the integrals diverge in $\omega = 0$, so the density of states $g(\omega)$ should have a different dependence on $\omega$ other than quadratic, to prevent the divergence. It is important to underline that the coupling between the tagged particle and the heat bath's degrees of freedom has been assumed to be constant with frequency. The divergence of these integrals may also be avoided by requiring the coupling to be a suitable function of $\omega$.

In the case of high external shear force, the process is expected to be generally non-Markovian because, as explained in the previous sections, the FDT has a linear combination of products of hyperbolic functions and trigonometric functions and this cannot be reduced to a Dirac delta function. Therefore, we expect that it won't be possible to recover a Markovian FDT for the Langevin equation with shear flow apart from the very special limit of nearly-vanishing shear rate and equilibrium initial conditions.

\section{Conclusion}

In the present work we have analytically derived the Langevin equation with shear flow and its corresponding fluctuation-dissipation theorems from first principles, for the first time. We followed a particle-bath Hamiltonian approach using a Caldeira-Leggett model supplemented with a term that describes the externally-applied mechanical deformation. The only microscopic Hamiltonian that describes a fluid flow is Hoover's DOLLS tensor Hamiltonian, which therefore has been chosen. For simplicity, but without loss of generality, we examined the easiest example of shear flow, that is planar Couette flow. We conclude that, in general, the model we proposed recovers the most important features of the Langevin models with shear used in the literature, but with some important differences. In particular, the Markovian limit is recovered only in the case of extremely low shear-rate values compared to the bath's thermal frequencies. We also demonstrated that it would be important to specify whether the initial state is in thermodynamic equilibrium or not, as this can lead to a different form of the fluctuation-dissipation relation. 
In a future perspective, it would be interesting to test the generalized non-Markovian Langevin equation with shear flow, that we derived here in its different limits, with numerical simulations. Moreover, this simple model may be used as the starting point to develop a new theory of the effective vibrational temperature in nonequilibrium sheared systems.

In future work, it could be useful to attempt a similar derivation as in \cite{McPhie} using the Mori-Zwanzig projection operator method to derive the GLE but crucially including also the derivation of the memory kernel. This could be done, e.g., by taking advantage of very recent advances in the field of memory kernel-reconstruction techniques for systems out of equilibrium \cite{Schilling_2022,Jung_2022}.
\\

\newpage
\section{Appendix}

\subsection{Compact form of Eq. \eqref{lang_small}}

We show how it is possible to write the generalized Langevin equation (GLE) Eq.\eqref{lang_small} in a compact form. If we group all the friction functions and the memory function we obtain 

\begin{multline}
 \dot{\vect{P}} = -V'[\vect{Q}]  -  \int_0^t \vect{K}(t-t') \dot{\vect{Q}}(t')
 \\+ \vect{F}(t) - \dot{\gamma} P_x(t) \hat{\vect{y}} + o(\dot{\gamma}^2)
 \end{multline}

\noindent with a $2 \times 2$ friction matrix

\begin{widetext}

\begin{equation}
    \vect{K}(t-t') = \begin{pmatrix} \begin{split}
        &\sum_i \frac{c_i^2}{\omega_i^2}\Bigl[ \cos\Bigl(\Bigl( \omega_i - \frac{\dot{\gamma}^2}{8\omega_i}\Bigr) (t-t') \Bigr)\cdot \\&\cos\Bigl(\frac{\dot{\gamma}}{2}(t-t')\Bigr)   \Bigr]  \end{split}  &\begin{split} &\sum_i \Bigl[\frac{c_i^2}{2\omega_i^2}(t-t') \cos\Bigl( \Bigl(\omega_i - \frac{\dot{\gamma}}{8\omega_i} \Bigr) (t-t')\Bigr)  \\&-  \frac{c_i^2}{2\omega_i^3} \sin\Bigl(\Bigl( \omega_i -\frac{\dot{\gamma}^2}{8\omega_i} \Bigr) (t-t') \Bigr)\Bigr]  \end{split}\\ \begin{split}& \sum_i \Bigl[-\frac{c_i^2}{2\omega_i^2} \cos\Bigl( \Bigl(\omega_i - \frac{\dot{\gamma}}{8\omega_i} \Bigr) (t-t')\Bigr) (t-t')  \\&+  \frac{c_i^2}{2\omega_i^3} \sin\Bigl(\Bigl( \omega_i - \frac{\dot{\gamma}^2}{8\omega_i} \Bigr) (t-t') \Bigr)\Bigr]  \end{split}
    & \begin{split}
        &\sum_i \Bigl[\frac{c_i^2}{\omega_i^2} \cos\Bigl(\Bigl( \omega_i + \frac{\dot{\gamma}^2}{8\omega_i}\Bigr) (t-t') \Bigr)\cos\Bigl(\frac{\dot{\gamma}}{2}(t-t')\Bigr)    \\&+ \frac{c_i^2}{\omega_i^4} \cos\left(\Bigl( \omega_i - \frac{\dot{\gamma}^2}{8\omega_i} \Bigr) (t-t') \right) \\& + \frac{c_i^2}{2\omega_i^3}(t-t') \sin\left(\Bigl( \omega_i - \frac{\dot{\gamma}^2}{8\omega_i} \Bigr) (t-t') \right) \Bigr]\end{split} 
\end{pmatrix}. 
\end{equation}

\end{widetext}

\subsection{Fluctuation-dissipation theorems}
In this section we present the details of the derivation of the fluctuation dissipation theorem (FDT). 

The main idea is to compute $\langle F_{x}(t) F_{x}(t') \rangle$ and $\langle F_{y}(t) F_{y}(t') \rangle$ separately, as they have different expressions, and then collect the components into a vector expression. 

\subsubsection{Equilibrium initial conditions}
If we take equilibrium initial conditions, then some terms in  $\langle F_{x}(t) F_{x}(t')\rangle$ and $\langle F_{y}(t) F_{y}(t') \rangle$ 
have their Boltzmann's ensemble average equal to zero: they are the mean of the Gaussian distribution given by $e^{-\Ham_B/k_B T}$. This Gaussian distribution is centered in $\vect{p}_i$ and $\Bigl(\vect{q}_i-c_i\frac{\vect{Q}}{\omega_i^2}\Bigr)$ therefore:

\[
 \langle p_x(0) \rangle =    \langle p_y(0) \rangle = 0 
\]
\[
\Bigl \langle \Bigl(\frac{c_i}{4}q_x(0) - \frac{c_i^2}{4\omega_i^2} Q_x(0) \Bigr) \Bigr \rangle = \Bigl \langle \Bigl(\frac{c_i}{4}q_y(0) - \frac{c_i^2}{4\omega_i^2} Q_y(0) \Bigr)\Bigr \rangle = 0
\]

\noindent 
Only the products between the other terms will contribute to the time-correlation function leading to 

\begin{widetext}

\begin{multline}
    \Bigl \langle \Bigl( \frac{c_i^2}{4\omega_i\sqrt{\omega_i(\dot{\gamma}-\omega_i)}}Q_y(0) - \frac{c_i\sqrt{\omega_i(\dot{\gamma}-\omega_i})}{4\omega_i}q_y(0)\Bigr)^2\Bigr \rangle = \frac{c_i^2(k_B T \omega_i^2(\dot{\gamma}-\omega_i)^2-c_i^2 Q_y(0)^2(\dot{\gamma}-2\omega_i)^2)}{16\omega_i^5(\dot{\gamma}-\omega_i)} 
    \end{multline}

\begin{multline}
\Bigl \langle\Bigl(-\frac{c_i^2}{4\omega_i\sqrt{\omega_i(\dot{\gamma}-\omega_i)}}Q_y(0) +\frac{c_i\sqrt{\omega_i(\dot{\gamma}-\omega_i})}{4\omega_i}q_y(0)\Bigr)^2\Bigr \rangle 
=-\frac{c_i^2(k_B T \omega_i^2(\dot{\gamma}+\omega_i)^2+c_i^2 Q_y^2(0) (\dot{\gamma}+2\omega_i)^2)}{16\omega_i^5(\dot{\gamma}+\omega_i)} 
\end{multline}

\begin{multline}
 \Bigl \langle \Bigl(
\frac{c_i^2}{4\omega_i\sqrt{\omega_i(\dot{\gamma}-\omega_i)}}Q_y(0) - \frac{c_i\sqrt{\omega_i(\dot{\gamma}-\omega_i})}{4\omega_i}q_y(0)\Bigr)\Bigl(-\frac{c_i^2}{4\omega_i\sqrt{\omega_i(\dot{\gamma}-\omega_i)}}Q_y(0) +\frac{c_i\sqrt{\omega_i(\dot{\gamma}-\omega_i})}{4\omega_i}q_y(0)\Bigr) \Bigr \rangle 
= \\ -\frac{c_i^2(k_B T \omega_i^2(\dot{\gamma}^2 -
\omega_i^2)
+c_i^2 Q_y^2(0) (\dot{\gamma}^2-4
\omega_i^2)}{16\omega_i^5\sqrt{\omega_i(\dot{\gamma}-\omega_i)}\sqrt{-\omega_i(\dot{\gamma}+\omega_i)}} 
\end{multline}

\begin{equation}
    \begin{split}
    &\Bigl \langle\Bigl[-\frac{c_i^2}{\omega_i^2}Q_x(0)+\frac{c_i}{4}q_x(0)+\frac{c_i}{4\omega_i} p_y(0)\Bigr]^2\Bigr \rangle = \frac{c_i^2}{8\omega_i^2}k_B T \\
    &\Bigl \langle\Bigl[-\frac{c_i^2}{\omega_i^2}Q_x(0)+\frac{c_i}{4}q_x(0)-\frac{c_i}{4\omega_i} p_y(0)\Bigr]^2\Bigr\rangle  = \frac{c_i^2}{8\omega_i^2}k_B T \\
& \langle{p}_x^2\rangle = \langle{p}_y^2\rangle = k_B T .\\  
    \end{split}
\end{equation}

\noindent in $\langle F_{x}(t) F_{x}(t') \rangle$ and to

\begin{multline}
    \Bigl \langle\Biggl(\frac{-c_i^2 Q_y(0)}{4\omega_i(\dot{\gamma}+\omega_i)}+ \frac{c_i}{4} q_y(0) -\frac{c_i}{4\omega_i} p_x\Biggr)^2 \Bigr \rangle  = \frac{-c_i^2(4 c_i^2 Q_y(0)^2 + k_B T \omega_i^2)}{16\omega_i^4}+\dot{\gamma}\frac{c_i^2 k_B T}{16\omega_i^3}
    \end{multline}
    
    \begin{multline}
    \Bigl \langle\Biggl(\frac{c_i^2 Q_y(0)}{4\omega_i(\dot{\gamma}-\omega_i)} + \frac{c_i}{4} q_y(0) +\frac{c_i}{4\omega_i} p_x\Biggr)^2 \Bigr \rangle =\frac{-c_i^2(4 c_i^2 Q_y(0)^2 + k_B T \omega_i^2)}{16\omega_i^4}-\dot{\gamma}\frac{c_i^2 k_B T}{16\omega_i^3}
    \end{multline}
    
    \begin{equation}
        \begin{split}
&\Bigl \langle\Biggl(\frac{c_i^2 Q_y(0)}{4\omega_i(\dot{\gamma}-\omega_i)} + \frac{c_i}{4} q_y(0) +\frac{c_i}{4\omega_i} p_x\Biggr) \Biggl(\frac{-c_i^2 Q_y(0)}{4\omega_i(\dot{\gamma}+\omega_i)}+ \frac{c_i}{4} q_y(0) -\frac{c_i}{4\omega_i} p_x\Biggr) \Bigr \rangle = \frac{c_i^2(4 c_i^2 Q_y(0)^2+k_B T \omega_i^2)}{16\omega_i^4}
\end{split}
    \end{equation}

\end{widetext}

in $\langle F_{y}(t) F_{y}(t') \rangle$. 

If a Taylor expansion is performed and if all the terms proportional to the tagged particle's initial position are neglected, as, with no loss of generality one could assume the tagged particle to be in the origin at $t=0$, then the expressions to be evaluated are:

\begin{widetext}
\begin{equation}\label{eq:noshear_iniz_x}
\begin{split} 
&\langle F_{x}(t) F_{x}(t') \rangle =\sum_i \Biggl\{ 
\Bigl[ -\frac{c_i^2  }{16\omega_i^2}k_B T+\dot{\gamma}\frac{c_i^2 }{16\omega_i^3} k_B T\Bigr] \Bigl(e^{\lambda_2 t}-e^{-\lambda_2 t}\Bigr) \Bigl(e^{\lambda_2 t'}-e^{-\lambda_2 t'}\Bigr)\\
&\Bigl[\frac{c_i^2 }{16\omega_i^2}k_B T+\dot{\gamma}\frac{c_i^2 }{16\omega_i^3} k_B T\Bigr]  \Bigl(e^{\lambda_4 t}-e^{-\lambda_4 t'}\Bigr)\Bigl(e^{\lambda_4 t'}-e^{-\lambda_4 t'}\Bigr)  +\frac{c_i^2 }{16\omega_i^2}k_B T 
\Bigl(e^{\lambda_2 t}-e^{-\lambda_2 t}\Bigr)(e^{\lambda_4 t'}-e^{-\lambda_4 t'}\Bigr) \\
&  +\frac{c_i^2 }{16\omega_i^4} k_B T
\Bigl(e^{\lambda_4 t}-e^{-\lambda_4 t}\Bigr)\Bigl(e^{\lambda_2 t}-e^{-\lambda_2 t}\Bigr)+ \frac{c_i^2}{8\omega_i^2} k_B T\Bigl(e^{\lambda_4 t}+e^{-\lambda_4 t}\Bigr)\Bigl(e^{\lambda_4 t'}+e^{-\lambda_4 t'}\Bigr) \\
&+  \frac{c_i^2}{8\omega_i^2} k_B T\Bigl(e^{\lambda_2 t}+e^{-\lambda_2 t}\Bigr)\Bigl(e^{\lambda_2 t'}+e^{-\lambda_2 t'}\Bigr) +\Bigl[-\frac{c_i^2}{16\omega_i^2}+ \dot{\gamma}\frac{c_i^2}{16\omega_i^3}\Bigr]
\Bigl(e^{-\lambda_2 t}-e^{\lambda_2 t}\Bigr)\Bigl(e^{-\lambda_2 t'}-e^{\lambda_2 t'}\Bigr) \\
&-  \frac{c_i^2}{16\omega_i^2}  k_B T  \Bigl(e^{-\lambda_2 t}+e^{-\lambda_2 t}\Bigr)e^{-\lambda_4 t'}+\Bigl[-\frac{c_i^2}{16\omega_i^2}+ \dot{\gamma}\frac{c_i^2}{16\omega_i^3}\Bigr]
\Bigl(e^{-\lambda_2 t}-e^{\lambda_2 t}\Bigr)\Bigl(e^{-\lambda_2 t'}-e^{\lambda_2 t'}\Bigr) \\
&-  \frac{c_i^2}{16\omega_i^2}  k_B T  \Bigl(e^{-\lambda_2 t}-e^{\lambda_2 t}\Bigr)e^{-\lambda_4 t'} +  \frac{c_i^2}{16\omega_i^2} k_B T \Bigl(e^{-\lambda_2 t}-e^{\lambda_2 t}\Bigr)e^{\lambda_4 t'} \\
&-\frac{c_i^2}{16\omega_i^2} k_B T 
e^{-\lambda_4 t}\Bigl(e^{-\lambda_2 t'}-e^{\lambda_2 t'}\Bigr)  +  \frac{c_i^2}{16\omega_i^2}  k_B T  e^{\lambda_4 t}\Bigl(e^{-\lambda_2 t'}-e^{\lambda_2 t'}\Bigr)\\
&+ \frac{c_i^2(\dot{\gamma}+\omega_i)}{16\omega_i^3} k_B T \Bigl(e^{-\lambda_4(t-t')}+e^{-\lambda_4(t-t')}-e^{-\lambda_4(t+t')}-e^{\lambda_4(t+t')}\Bigr)\\
\end{split}
\end{equation}

\noindent and

\begin{equation} \label{eq:noshear_iniz_y}
    \begin{split}
      &\langle F_{y}(t) F_{y}(t') \rangle   =\sum_i\Biggl\{ \frac{c_i^2}{8\omega_i^2} \Bigl(e^{\lambda_2 t}+e^{-\lambda_2 t}\Bigr)\Bigl(e^{\lambda_2 t'}+e^{-\lambda_2 t'}\Bigr)  +\frac{c_i^2}{8\omega_i^4} \Bigl(e^{\lambda_4 t}+e^{-\lambda_4 t}\Bigr)\Bigl(e^{\lambda_4 t'}+e^{-\lambda_4 t'}\Bigr) \\
      & - \Bigl[\frac{c_i^2}{8\omega_i^2} + \dot{\gamma} \frac{c_i^2}{8\omega_i^3}\Bigr] \Bigl(e^{\lambda_2 t}-e^{-\lambda_2 t}\Bigr)\Bigl(e^{\lambda_2 t'}-e^{-\lambda_2 t'}\Bigr) +\Bigl[-\frac{c_i^2}{8\omega_i^2} + \dot{\gamma} \frac{c_i^2}{8\omega_i^3}\Bigr] \Bigl(e^{\lambda_4 t}-e^{-\lambda_4 t}\Bigr)\Bigl(e^{\lambda_4 t'}-e^{-\lambda_4 t'}\Bigr) \Biggr\}.
    \end{split}
\end{equation}

As we are in the low shear-rate regime, then the eigenvalues become

\begin{equation}
    \begin{split}
        \lambda_2 = i \Biggl(\omega_i -\frac{\dot{\gamma}}{2}-\frac{\dot{\gamma}^2}{8\omega_i}\Biggr) \\
        \lambda_4 = i \Biggl(\omega_i +\frac{\dot{\gamma}}{2}-\frac{\dot{\gamma}^2}{8\omega_i}\Biggr) \\
    \end{split}
\end{equation}

\noindent and the exponential factors then read as:

\begin{equation}
\begin{split}
   & e^{\lambda_2 t }- e^{-\lambda_2 t} = 2i\sin\Biggl(\Biggl(\omega_i -\frac{\dot{\gamma}}{2}-\frac{\dot{\gamma}^2}{8\omega_i}\Biggr)t\Biggr) \\
    & e^{\lambda_4 t }- e^{-\lambda_4 t} = 2i\sin\Biggl(\Biggl(\omega_i +\frac{\dot{\gamma}}{2}-\frac{\dot{\gamma}^2}{8\omega_i}\Biggr)t\Biggr) \\
     & e^{\lambda_2 t }+ e^{-\lambda_2 t} = 2\cos\Biggl(\Biggl(\omega_i -\frac{\dot{\gamma}}{2}-\frac{\dot{\gamma}^2}{8\omega_i}\Biggr)t\Biggr) \\
      & e^{\lambda_4 t }+ e^{-\lambda_4t} = 2\cos\Biggl(\Biggl(\omega_i +\frac{\dot{\gamma}}{2}-\frac{\dot{\gamma}^2}{8\omega_i}\Biggr)t\Biggr). \\
    \end{split}
\end{equation}

By grouping the terms having the same coefficients, and performing some algebric manipulations using standard trigonometric identities, then  Eq. \eqref{FDT_eq} is found.

\subsubsection{Nonequilibrium initial conditions}

In this case the Boltzmann averages are:

\begin{multline}\label{eq:noneq_x1}
    \Bigl \langle \Bigl( \frac{c_i^2}{4\omega_i\sqrt{\omega_i(\dot{\gamma}-\omega_i)}}Q_y(0) - \frac{c_i\sqrt{\omega_i(\dot{\gamma}-\omega_i})}{4\omega_i}q_y(0)\Bigr)^2\Bigr \rangle = \frac{-c_i^2 k_B T }{16\omega_i^2}+\dot{\gamma}\frac{c_i^2 k_B T}{16\omega_i^3} +\dot{\gamma}^2 \frac{c_i^2}{4\omega_i^4}k_B T
    \end{multline}

\begin{multline}\label{eq:noneq_x2}
\Bigl \langle\Bigl(-\frac{c_i^2}{4\omega_i\sqrt{-\omega_i(\dot{\gamma}+\omega_i)}}Q_y(0) +\frac{c_i\sqrt{-\omega_i(\dot{\gamma}+\omega_i})}{4\omega_i}q_y(0)\Bigr)^2\Bigr \rangle 
=\frac{-c_i^2 k_B T }{16\omega_i^2}-\dot{\gamma}\frac{c_i^2 k_B T}{16\omega_i^3}+\dot{\gamma}^2 \frac{c_i^2}{4\omega_i^4}k_B T
\end{multline}

\begin{multline}\label{eq:noneq_x3}
 \Bigl \langle \Bigl(
\frac{c_i^2}{4\omega_i\sqrt{\omega_i(\dot{\gamma}-\omega_i)}}Q_y(0) - \frac{c_i\sqrt{\omega_i(\dot{\gamma}-\omega_i})}{4\omega_i}q_y(0)\Bigr)\Bigl(-\frac{c_i^2}{4\omega_i\sqrt{\omega_i(\dot{\gamma}-\omega_i)}}Q_y(0) +\frac{c_i\sqrt{\omega_i(\dot{\gamma}-\omega_i})}{4\omega_i}q_y(0)\Bigr) \Bigr \rangle 
\\=  \frac{c_i^2 k_B T }{16\omega_i^2} + \dot{\gamma}^2 \frac{7 c_i^2}{32\omega_i^4}k_B T
\end{multline}

\begin{equation}
\label{eq:noneq_x4}
 \langle p^2_x(0) \rangle = k_B T + \frac{4\dot{\gamma}^2 }{\omega_i^4}   
\end{equation}

\begin{multline}\label{eq:noneq_x5}
\Bigl \langle  \Bigl(\frac{c_i^2}{4\omega_i\sqrt{\omega_i(\dot{\gamma}+\omega_i)}}Q_y(0) -\frac{c_i\sqrt{\omega_i(\dot{\gamma}+\omega_i})}{4\omega_i}q_y(0)\Bigr) p_x(0)  \Bigr \rangle  =\dot{\gamma}\frac{c_i}{2 i \omega_i^2}
\end{multline}

\begin{multline}\label{eq:noneq_x6}
\Bigl \langle  \Bigl(-\frac{c_i^2}{4\omega_i\sqrt{\omega_i(\dot{\gamma}-\omega_i)}}Q_y(0) +\frac{c_i\sqrt{\omega_i(\dot{\gamma}+\omega_i})}{4\omega_i}q_y(0)\Bigr) p_x(0)  \Bigr \rangle =
-\dot{\gamma}\frac{c_i}{2 i \omega_i^2}
\end{multline}

and

\begin{multline} \label{eq:noneq_y_1}
\Bigl \langle\Biggl(\frac{-c_i^2 Q_y(0)}{4\omega_i(\dot{\gamma}+\omega_i)}+ \frac{c_i}{4} q_y(0) -\frac{c_i}{4\omega_i} p_x\Biggr)^2 \Bigr \rangle = \frac{c_i^2}{8\omega_i^2}  +\dot{\gamma}\frac{c_i^2}{4\omega_i^3}k_B T +\dot{\gamma}^2 \frac{c_i^2}{2\omega_i^4}k_B T  
\end{multline}

\begin{multline} \label{eq:noneq_y_2}
\Bigl \langle\Biggl(\frac{c_i^2 Q_y(0)}{4\omega_i(\dot{\gamma}-\omega_i} + \frac{c_i}{4} q_y(0) +\frac{c_i}{4\omega_i} p_x\Biggr)^2 \Bigr \rangle =\frac{c_i^2}{8\omega_i^2}  - \dot{\gamma}\frac{c_i^2}{4\omega_i^3}k_B T +\dot{\gamma}^2 \frac{c_i^2}{2\omega_i^4}k_B T  
\end{multline}

\begin{multline}\label{eq:noneq_y_3}
 \Bigl \langle \Biggl(\frac{c_i^2}{4\omega_i\sqrt{-\omega_i(\dot{\gamma}+\omega_i)}} - \frac{c_i\omega_i q_x(0)}{4\sqrt{-\omega_i(\dot{\gamma}+\omega_i)}}    +\frac{c_i}{\sqrt{-\omega_i(\dot{\gamma}+\omega_i)}}p_y(0)\Biggr)^2 \Bigr\rangle  = -\frac{c_i}{8\omega_i^2} +\dot{\gamma}\frac{c_i}{8\omega_i^3} - \dot{\gamma}^2  \frac{c_i}{8\omega_i^4}
\end{multline}

\begin{multline}\label{eq:noneq_y_4}
 \Bigl \langle \Biggl(-\frac{c_i^2}{4\omega_i\sqrt{\omega_i(\dot{\gamma}-\omega_i)}} + \frac{c_i\omega_i q_x(0)}{4\sqrt{\omega_i(\dot{\gamma}-\omega_i)}}   +\frac{c_i}{\sqrt{-\omega_i(\dot{\gamma}+\omega_i)}}p_y(0)\Biggr)^2 \Bigr\rangle   = -\frac{c_i}{8\omega_i^2} -\dot{\gamma}\frac{c_i}{8\omega_i^3} - \dot{\gamma}^2  \frac{c_i}{8\omega_i^4}
\end{multline}

\noindent the other products do not contribute as their average is proportional to the tagged particle's initial conditions. 

As in this case there are some additional terms with nonzero Boltzmann average, then it is possible to take the FDT computed with equilibrium initial conditions and then just add the new terms.

We first take the $x$ component and the terms in Eqs. (\ref{eq:noneq_x1})-(\ref{eq:noneq_x3}) which give

\begin{equation}
    \begin{split}
        &\frac{c_i^2}{4\omega_i^2} k_B T \Bigl[ \sin\Bigl(\Bigl((\omega_i -\frac{\dot{\gamma}}{2} -\frac{\dot{\gamma}^2}{8\omega_i}\Bigr)t\Bigr) \sin\Bigl(\Bigl(\omega_i -\frac{\dot{\gamma}}{2} -\frac{\dot{\gamma}^2}{8\omega_i}\Bigr)t'\Bigr) +\sin\Bigl(\Bigl((\omega_i +\frac{\dot{\gamma}}{2} -\frac{\dot{\gamma}^2}{8\omega_i}\Bigr)t\Bigr) \sin\Bigl(\Bigl(\omega_i +\frac{\dot{\gamma}}{2} -\frac{\dot{\gamma}^2}{8\omega_i}\Bigr)t'\Bigr)  \\
        &- \sin\Bigl(\Bigl((\omega_i -\frac{\dot{\gamma}}{2} -\frac{\dot{\gamma}^2}{8\omega_i}\Bigr)t\Bigr) \sin\Bigl(\Bigl(\omega_i +\frac{\dot{\gamma}}{2}-\frac{\dot{\gamma}^2}{8\omega_i}\Bigr)t'\Bigr)-\sin\Bigl(\Bigl((\omega_i +\frac{\dot{\gamma}}{2} -\frac{\dot{\gamma}^2}{8\omega_i}\Bigr)t\Bigr) \sin\Bigl(\Bigl(\omega_i -\frac{\dot{\gamma}}{2}-\frac{\dot{\gamma}^2}{8\omega_i}\Bigr)t'\Bigr)  \Bigr]\\
        &+\dot{\gamma}\frac{c_i^2}{4\omega_i^2}k_B T \Bigl[ \sin\Bigl(\Bigl((\omega_i +\frac{\dot{\gamma}}{2} -\frac{\dot{\gamma}^2}{8\omega_i}\Bigr)t\Bigr) \sin\Bigl(\Bigl(\omega_i +\frac{\dot{\gamma}}{2} -\frac{\dot{\gamma}^2}{8\omega_i}\Bigr)t'\Bigr) -\sin\Bigl(\Bigl((\omega_i -\frac{\dot{\gamma}}{2} -\frac{\dot{\gamma}^2}{8\omega_i}\Bigr)t\Bigr)\\
        & \sin\Bigl(\Bigl(\omega_i -\frac{\dot{\gamma}}{2} -\frac{\dot{\gamma}^2}{8\omega_i}\Bigr)t'\Bigr) \Bigr] -\dot{\gamma}^2 \frac{c_i^2}{4\omega_i^4} k_B T \Bigl[ \sin\Bigl(\Bigl((\omega_i -\frac{\dot{\gamma}}{2}-\frac{\dot{\gamma}^2}{8\omega_i}\Bigr)t\Bigr) \sin\Bigl(\Bigl(\omega_i -\frac{\dot{\gamma}}{2} -\frac{\dot{\gamma}^2}{8\omega_i}\Bigr)t'\Bigr)\\
        &+ \sin\Bigl(\Bigl((\omega_i +\frac{\dot{\gamma}}{2} -\frac{\dot{\gamma}^2}{8\omega_i}\Bigr)t\Bigr) \sin\Bigl(\Bigl(\omega_i +\frac{\dot{\gamma}}{2} -\frac{\dot{\gamma}^2}{8\omega_i}\Bigr)t'\Bigr)  +\frac{7}{8} \sin\Bigl(\Bigl((\omega_i -\frac{\dot{\gamma}}{2} -\frac{\dot{\gamma}^2}{8\omega_i}\Bigr)t\Bigr) \sin\Bigl(\Bigl(\omega_i +\frac{\dot{\gamma}}{2} -\frac{\dot{\gamma}^2}{8\omega_i}\Bigr)t'\Bigr) \\
        &+ \frac{7}{8} \sin\Bigl(\Bigl((\omega_i +\frac{\dot{\gamma}}{2} -\frac{\dot{\gamma}^2}{8\omega_i}\Bigr)t\Bigr) \sin\Bigl(\Bigl(\omega_i -\frac{\dot{\gamma}}{2} -\frac{\dot{\gamma}^2}{8\omega_i}\Bigr)t'\Bigr) \Bigr] 
    \end{split}
\end{equation}

Then we analyze the extra term coming from $\langle p_x^2(0)\rangle$ in Eq.(\ref{eq:noneq_x4}) 
\begin{equation}
    \begin{split}
    & \dot{\gamma}^2 \frac{c_i^2}{\omega_i^2}k_B T\Bigl[ \sin\Bigl(\Bigl(\omega_i -\frac{\dot{\gamma}}{2} -\frac{\dot{\gamma}^2}{8\omega_i}\Bigr)t\Bigr) \sin\Bigl(\Bigl(\omega_i +\frac{\dot{\gamma}}{2} -\frac{\dot{\gamma}^2}{8\omega_i}\Bigr)t'\Bigr)\Bigr) + \sin\Bigl(\Bigl((\omega_i +\frac{\dot{\gamma}}{2} -\frac{\dot{\gamma}^2}{8\omega_i}\Bigr)t\Bigr) \sin\Bigl(\Bigl(\omega_i -\frac{\dot{\gamma}}{2} -\frac{\dot{\gamma}^2}{8\omega_i}\Bigr)t'\Bigr)\Bigr) \\
     &+ \sin\Bigl(\Bigl((\omega_i -\frac{\dot{\gamma}}{2} -\frac{\dot{\gamma}^2}{8\omega_i}\Bigr)t\Bigr) \sin\Bigl(\Bigl(\omega_i -\frac{\dot{\gamma}}{2} -\frac{\dot{\gamma}^2}{8\omega_i}\Bigr)t'\Bigr)) + \sin\Bigl(\Bigl((\omega_i +\frac{\dot{\gamma}}{2} -\frac{\dot{\gamma}^2}{8\omega_i}\Bigr)t\Bigr) \sin\Bigl(\Bigl(\omega_i +\frac{\dot{\gamma}}{2} -\frac{\dot{\gamma}^2}{8\omega_i}\Bigr)t'\Bigr) \Bigr]\\
    \end{split}
\end{equation}

and from the mixed products in Eqs.(\ref{eq:noneq_x5})-(\ref{eq:noneq_x6}) that give
\begin{equation}
    \begin{split}
        &\frac{c_i^2}{\omega_i^2} k_B T \Bigl[\sin\Bigl(\Bigl((\omega_i +\frac{\dot{\gamma}}{2} -\frac{\dot{\gamma}^2}{8\omega_i}\Bigr)t\Bigr) \sin\Bigl(\Bigl(\omega_i +\frac{\dot{\gamma}}{2} -\frac{\dot{\gamma}^2}{8\omega_i}\Bigr)t'\Bigr)- \sin\Bigl(\Bigl((\omega_i -\frac{\dot{\gamma}}{2} -\frac{\dot{\gamma}^2}{8\omega_i}\Bigr)t\Bigr) \sin\Bigl(\Bigl(\omega_i -\frac{\dot{\gamma}}{2} -\frac{\dot{\gamma}^2}{8\omega_i}\Bigr)t'\Bigr) \Bigr] \\
        & + \dot{\gamma}\frac{c_i^2}{2\omega_i^3} k_B T \Bigl[\sin\Bigl(\Bigl((\omega_i -\frac{\dot{\gamma}}{2} -\frac{\dot{\gamma}^2}{8\omega_i}\Bigr)t\Bigr) \sin\Bigl(\Bigl(\omega_i -\frac{\dot{\gamma}}{2} -\frac{\dot{\gamma}^2}{8\omega_i}\Bigr)t'\Bigr) -\sin\Bigl(\Bigl((\omega_i +\frac{\dot{\gamma}}{2} -\frac{\dot{\gamma}^2}{8\omega_i}\Bigr)t\Bigr) \sin\Bigl(\Bigl(\omega_i +\frac{\dot{\gamma}}{2} -\frac{\dot{\gamma}^2}{8\omega_i}\Bigr)t'\Bigr)
        \Bigr] \\&+ \dot{\gamma}^2 \frac{c_i^2}{8\omega_i^4}k_B T \Bigl[\sin\Bigl(\Bigl((\omega_i -\frac{\dot{\gamma}}{2} -\frac{\dot{\gamma}^2}{8\omega_i}\Bigr)t\Bigr) \sin\Bigl(\Bigl(\omega_i -\frac{\dot{\gamma}}{2} -\frac{\dot{\gamma}^2}{8\omega_i}\Bigr)t'\Bigr)-\sin\Bigl(\Bigl((\omega_i +\frac{\dot{\gamma}}{2} -\frac{\dot{\gamma}^2}{8\omega_i}\Bigr)t\Bigr) \sin\Bigl(\Bigl(\omega_i +\frac{\dot{\gamma}}{2} -\frac{\dot{\gamma}^2}{8\omega_i}\Bigr)t'\Bigr)
        \Bigr] .\\ 
    \end{split}
\end{equation}

In $\langle F_{y}(t) F_{y}(t') \rangle$ the  additional terms come from Eqs.(\ref{eq:noneq_y_1})-(\ref{eq:noneq_y_2}) :

\begin{equation}
    \begin{split}
       & \dot{\gamma}\frac{c_i^2}{\omega_i^3}k_B T \Bigl[ \cos\Bigl(\Bigl((\omega_i -\frac{\dot{\gamma}}{2} -\frac{\dot{\gamma}^2}{8\omega_i}\Bigr)t\Bigr) \cos\Bigl(\Bigl((\omega_i -\frac{\dot{\gamma}}{2} -\frac{\dot{\gamma}^2}{8\omega_i}\Bigr)t'\Bigr) -\cos\Bigl(\Bigl((\omega_i +\frac{\dot{\gamma}}{2} -\frac{\dot{\gamma}^2}{8\omega_i}\Bigr)t\Bigr) \cos\Bigl(\Bigl((\omega_i +\frac{\dot{\gamma}}{2} -\frac{\dot{\gamma}^2}{8\omega_i}\Bigr)t'\Bigr) \Bigr] \\
        &+ 2\dot{\gamma}^2\frac{c_i^2}{\omega_i^4}k_B T \Bigl[ \cos\Bigl(\Bigl((\omega_i -\frac{\dot{\gamma}}{2} -\frac{\dot{\gamma}^2}{8\omega_i}\Bigr)t\Bigr) \cos\Bigl(\Bigl((\omega_i -\frac{\dot{\gamma}}{2} -\frac{\dot{\gamma}^2}{8\omega_i}\Bigr)t'\Bigr) +\cos\Bigl(\Bigl((\omega_i +\frac{\dot{\gamma}}{2} -\frac{\dot{\gamma}^2}{8\omega_i}\Bigr)t\Bigr) \cos\Bigl(\Bigl((\omega_i +\frac{\dot{\gamma}}{2} -\frac{\dot{\gamma}^2}{8\omega_i}\Bigr)t'\Bigr) \Bigr]
    \end{split}
\end{equation}

and those from Eqs.(\ref{eq:noneq_y_3})-(\ref{eq:noneq_y_4}) are

\begin{equation}
    \begin{split}
      &  \frac{c_i^2}{2\omega_i^2}k_B T \Bigl[ \sin\Bigl(\Bigl((\omega_i -\frac{\dot{\gamma}}{2} -\frac{\dot{\gamma}^2}{8\omega_i}\Bigr)t\Bigr) \sin\Bigl(\Bigl((\omega_i -\frac{\dot{\gamma}}{2} -\frac{\dot{\gamma}^2}{8\omega_i}\Bigr)t'\Bigr)+ \sin\Bigl(\Bigl((\omega_i +\frac{\dot{\gamma}}{2} -\frac{\dot{\gamma}^2}{8\omega_i}\Bigr)t\Bigr) \sin\Bigl(\Bigl((\omega_i +\frac{\dot{\gamma}}{2} -\frac{\dot{\gamma}^2}{8\omega_i}\Bigr)t'\Bigr) \\
        & + \dot{\gamma}  \frac{c_i^2}{2\omega_i^3}k_B T \Bigl[ \sin\Bigl(\Bigl((\omega_i +\frac{\dot{\gamma}}{2} -\frac{\dot{\gamma}^2}{8\omega_i}\Bigr)t\Bigr) \sin\Bigl(\Bigl((\omega_i +\frac{\dot{\gamma}}{2} -\frac{\dot{\gamma}^2}{8\omega_i}\Bigr)t'\Bigr) - \sin\Bigl(\Bigl((\omega_i -\frac{\dot{\gamma}}{2} -\frac{\dot{\gamma}^2}{8\omega_i}\Bigr)t\Bigr) \sin\Bigl(\Bigl((\omega_i -\frac{\dot{\gamma}}{2} -\frac{\dot{\gamma}^2}{8\omega_i}\Bigr)t'\Bigr) \Bigr] \\
        &+ \dot{\gamma}^2 \frac{c_i^2}{2\omega_i^4}k_B T \Bigl[\sin\Bigl(\Bigl((\omega_i +\frac{\dot{\gamma}}{2} -\frac{\dot{\gamma}^2}{8\omega_i}\Bigr)t\Bigr) \sin\Bigl(\Bigl((\omega_i +\frac{\dot{\gamma}}{2} -\frac{\dot{\gamma}^2}{8\omega_i}\Bigr)t'\Bigr)- \sin\Bigl(\Bigl((\omega_i -\frac{\dot{\gamma}}{2} -\frac{\dot{\gamma}^2}{8\omega_i}\Bigr)t\Bigr) \sin\Bigl(\Bigl((\omega_i -\frac{\dot{\gamma}}{2} -\frac{\dot{\gamma}^2}{8\omega_i}\Bigr)t'\Bigr) \Bigr].  \\
    \end{split}
\end{equation}
  
By plugging these terms into Eq.\eqref{FDT_eq} and by performing standard trigonometric manipulations then Eq.\eqref{FDT_noneq} is found. 

\end{widetext}

\begin{acknowledgments}
This work has received funding from the European Union (ERC, ``Multimech'', contract no. 101043968) and from the US Army Research Office through contract nr.   W911NF-22-2-0256. 
\end{acknowledgments}



\bibliography{references}


\end{document}